\pdfoutput=1
\documentclass[12pt,a4paper]{article}
\usepackage{array}
\usepackage{ifthen} % for conditional statements
\newboolean{pdflatex}
\setboolean{pdflatex}{true} % False for eps figures 

\newboolean{articletitles}
\setboolean{articletitles}{true} % False removes titles in references

\newboolean{uprightparticles}
\setboolean{uprightparticles}{false} %True for upright particle symbols

\def\paperauthors{LHCb collaboration} % Leave as is for PAPER, CONF and FIGURE
\def\paperasciititle{Observation of an excited neutral Xib state} 
\def\papertitle{Observation of a new $\Xibz$ state}
\def\paperkeywords{{High Energy Physics}, {LHCb}} % Comma separated list
\def\papercopyright{\the\year\ CERN for the benefit of the LHCb collaboration} % new since 9/Apr/2018
\def\paperlicence{CC BY 4.0 licence}
\def\paperlicenceurl{https://creativecommons.org/licenses/by/4.0/}

\usepackage[top=1in, bottom=1.25in, left=1in, right=1in]{geometry}

\usepackage{microtype}
\usepackage{lineno}  % for line numbering during review
\usepackage{xspace} % To avoid problems with missing or double spaces after
\usepackage{caption} %these three command get the figure and table captions automatically small

\usepackage{graphicx}  % to include figures (can also use other packages)
\usepackage{color}
\usepackage{colortbl}
\graphicspath{{./figs/}} % Make Latex search fig subdir for figures
\usepackage{amsmath} % Adds a large collection of math symbols
\usepackage{amssymb}
\usepackage{amsfonts}
\usepackage{upgreek} % Adds in support for greek letters in roman typeset

\newcommand*\patchAmsMathEnvironmentForLineno[1]{%
\expandafter\let\csname old#1\expandafter\endcsname\csname #1\endcsname
\expandafter\let\csname oldend#1\expandafter\endcsname\csname
end#1\endcsname
 \renewenvironment{#1}%
   {\linenomath\csname old#1\endcsname}%
   {\csname oldend#1\endcsname\endlinenomath}%
}
\newcommand*\patchBothAmsMathEnvironmentsForLineno[1]{%
  \patchAmsMathEnvironmentForLineno{#1}%
  \patchAmsMathEnvironmentForLineno{#1*}%
}
\AtBeginDocument{%
\patchBothAmsMathEnvironmentsForLineno{equation}%
\patchBothAmsMathEnvironmentsForLineno{align}%
\patchBothAmsMathEnvironmentsForLineno{flalign}%
\patchBothAmsMathEnvironmentsForLineno{alignat}%
\patchBothAmsMathEnvironmentsForLineno{gather}%
\patchBothAmsMathEnvironmentsForLineno{multline}%
\patchBothAmsMathEnvironmentsForLineno{eqnarray}%
}

\usepackage{hyperxmp}

\usepackage[pdftex,
            pdfauthor={\paperauthors},
            pdftitle={\paperasciititle},
            pdfkeywords={\paperkeywords},
            pdfcopyright={Copyright (C) \papercopyright},
            pdflicenseurl={\paperlicenceurl}]{hyperref}
\usepackage[bottom,flushmargin,hang,multiple]{footmisc}

\usepackage[all]{hypcap} % Internal hyperlinks to floats.

\usepackage{xspace} 
\usepackage{upgreek}

\def\XibStarm         {{\ensuremath{\Xires_\bquark(6227)^{-}}}\xspace}
\def\XibStarz         {{\ensuremath{\Xires_\bquark(6227)^{0}}}\xspace}
\def\XicPrime         {{\ensuremath{\Xires_\cquark}^{\prime0}}\xspace}
\def\XibStarZero      {{\ensuremath{\Xires_\bquark(5945)^{0}}}\xspace}
\def\dmPiPeak         {\delta m^{\rm peak}_{\pi}}
\def\dmKPeak         {\delta m^{\rm peak}_{K}}

\def\lhcb   {\mbox{LHCb}\xspace}

\def\MagUp {\mbox{\em Mag\kern -0.05em Up}\xspace}

\ifthenelse{\boolean{uprightparticles}}%
{

 \def\Pmu         {\ensuremath{\upmu}\xspace}

 \def\Ppi         {\ensuremath{\uppi}\xspace}

 \def\Ppsi        {\ensuremath{\uppsi}\xspace}

 \def\PDelta      {\ensuremath{\Delta}\xspace}                 
 \def\PXi         {\ensuremath{\Xi}\xspace}                 
 \def\PLambda     {\ensuremath{\Lambda}\xspace}                 
 \def\PSigma      {\ensuremath{\Sigma}\xspace}                 
 \def\POmega      {\ensuremath{\Omega}\xspace}                 
 \def\PUpsilon    {\ensuremath{\Upsilon}\xspace}

 \def\PB      {\ensuremath{\mathrm{B}}\xspace}                 
                  
 \def\PD      {\ensuremath{\mathrm{D}}\xspace}

 \def\PJ      {\ensuremath{\mathrm{J}}\xspace}                 
 \def\PK      {\ensuremath{\mathrm{K}}\xspace}

 \def\Pb      {\ensuremath{\mathrm{b}}\xspace}                 
 \def\Pc      {\ensuremath{\mathrm{c}}\xspace}

 \def\Pi      {\ensuremath{\mathrm{i}}\xspace}

 \def\Ps      {\ensuremath{\mathrm{s}}\xspace}

 \def\thebaroffset{0.0em}
}
{

 \def\Pmu         {\ensuremath{\mu}\xspace}

 \def\Ppi         {\ensuremath{\pi}\xspace}

 \def\Ppsi        {\ensuremath{\psi}\xspace}                 
                  
 \mathchardef\PDelta="7101
 \mathchardef\PXi="7104
 \mathchardef\PLambda="7103
 \mathchardef\PSigma="7106
 \mathchardef\POmega="710A
 \mathchardef\PUpsilon="7107
                  
 \def\PB      {\ensuremath{B}\xspace}                 
                  
 \def\PD      {\ensuremath{D}\xspace}

 \def\PJ      {\ensuremath{J}\xspace}                 
 \def\PK      {\ensuremath{K}\xspace}

 \def\Pb      {\ensuremath{b}\xspace}                 
 \def\Pc      {\ensuremath{c}\xspace}

 \def\Pi      {\ensuremath{i}\xspace}

 \def\Ps      {\ensuremath{s}\xspace}

 \def\thebaroffset{0.18em}
}
\newcommand{\offsetoverline}[2][\thebaroffset]{\kern #1\overline{\kern -#1 #2}}%

\makeatletter
\ifcase \@ptsize \relax% 10pt
  \newcommand{\miniscule}{\@setfontsize\miniscule{4}{5}}% \tiny: 5/6
\or% 11pt
  \newcommand{\miniscule}{\@setfontsize\miniscule{5}{6}}% \tiny: 6/7
\or% 12pt
  \newcommand{\miniscule}{\@setfontsize\miniscule{5}{6}}% \tiny: 6/7
\fi
\makeatother

\DeclareRobustCommand{\optbar}[1]{\shortstack{{\miniscule (\rule[.5ex]{1.25em}{.18mm})}
  \\ [-.7ex] $#1$}}

\def\mup        {{\ensuremath{\Pmu^+}}\xspace}
\def\mun        {{\ensuremath{\Pmu^-}}\xspace} % muon negative (\mum is taken)

\def\squark    {{\ensuremath{\Ps}}\xspace}

\def\cquark    {{\ensuremath{\Pc}}\xspace}

\def\bquark    {{\ensuremath{\Pb}}\xspace}

\def\pion   {{\ensuremath{\Ppi}}\xspace}
\def\piz    {{\ensuremath{\pion^0}}\xspace}
\def\pip    {{\ensuremath{\pion^+}}\xspace}
\def\pim    {{\ensuremath{\pion^-}}\xspace}

\def\kaon    {{\ensuremath{\PK}}\xspace}
\def\Kbar    {{\ensuremath{\offsetoverline{\PK}}}\xspace}

\def\KorKbar {\kern \thebaroffset\optbar{\kern -\thebaroffset \PK}{}\xspace}

\def\Kzb     {{\ensuremath{\Kbar{}^0}}\xspace}
\def\Kp      {{\ensuremath{\kaon^+}}\xspace}
\def\Km      {{\ensuremath{\kaon^-}}\xspace}

\def\D       {{\ensuremath{\PD}}\xspace}

\def\DorDbar {\kern \thebaroffset\optbar{\kern -\thebaroffset \PD}\xspace}
\def\Dz      {{\ensuremath{\D^0}}\xspace}

\def\Dp      {{\ensuremath{\D^+}}\xspace}
\def\Dm      {{\ensuremath{\D^-}}\xspace}

\def\DpDm    {\ensuremath{\Dp {\kern -0.16em \Dm}}\xspace}

\def\Dstarp  {{\ensuremath{\D^{*+}}}\xspace}

\def\B       {{\ensuremath{\PB}}\xspace}

\def\BorBbar {\kern \thebaroffset\optbar{\kern -\thebaroffset \PB}\xspace}

\def\Bd      {{\ensuremath{\B^0}}\xspace}

\def\BdorBdbar {\kern \thebaroffset\optbar{\kern -\thebaroffset \Bd}\xspace}

\def\Bs      {{\ensuremath{\B^0_\squark}}\xspace}

\def\BsorBsbar {\kern \thebaroffset\optbar{\kern -\thebaroffset \Bs}\xspace}

\def\jpsi     {{\ensuremath{{\PJ\mskip -3mu/\mskip -2mu\Ppsi}}}\xspace}

\def\Y#1S{\ensuremath{\PUpsilon{(#1S)}}\xspace}

\def\Lz          {{\ensuremath{\PLambda}}\xspace}

\def\LorLbar     {\kern \thebaroffset\optbar{\kern -\thebaroffset \PLambda}\xspace}

\def\Xires       {{\ensuremath{\PXi}}\xspace}

\def\Lc          {{\ensuremath{\Lz^+_\cquark}}\xspace}

\def\Xicz        {{\ensuremath{\Xires^0_\cquark}}\xspace}

\def\Lb           {{\ensuremath{\Lz^0_\bquark}}\xspace}

\def\Xib          {{\ensuremath{\Xires_\bquark}}\xspace}
\def\Xibz         {{\ensuremath{\Xires^0_\bquark}}\xspace}
\def\Xibm         {{\ensuremath{\Xires^-_\bquark}}\xspace}

\def\BF         {{\ensuremath{\mathcal{B}}}\xspace}
\def\BR         {\BF}

\def\to                 {\ensuremath{\rightarrow}\xspace}

\def\AT#1     {\ensuremath{A_{\mathrm{T}}^{#1}}\xspace}           % 2

\def\C#1      {\ensuremath{\mathcal{C}_{#1}}\xspace}                       % 9
\def\Cp#1     {\ensuremath{\mathcal{C}_{#1}^{'}}\xspace}                    % 7
\def\Ceff#1   {\ensuremath{\mathcal{C}_{#1}^{\mathrm{(eff)}}}\xspace}        % 9  
\def\Cpeff#1  {\ensuremath{\mathcal{C}_{#1}^{'\mathrm{(eff)}}}\xspace}       % 7
\def\Ope#1    {\ensuremath{\mathcal{O}_{#1}}\xspace}                       % 2
\def\Opep#1   {\ensuremath{\mathcal{O}_{#1}^{'}}\xspace}                    % 7

\newcommand{\nospaceunit}[1]{\ensuremath{\text{#1}}}       
\newcommand{\aunit}[1]{\ensuremath{\text{\,#1}}}       
\newcommand{\tev}{\aunit{Te\kern -0.1em V}\xspace}
\newcommand{\gev}{\aunit{Ge\kern -0.1em V}\xspace}
\newcommand{\mev}{\aunit{Me\kern -0.1em V}\xspace}
\newcommand{\kev}{\aunit{ke\kern -0.1em V}\xspace}
\newcommand{\ev}{\aunit{e\kern -0.1em V}\xspace}
 
\newcommand{\mevc}{\ensuremath{\aunit{Me\kern -0.1em V\!/}c}\xspace}
\newcommand{\gevc}{\ensuremath{\aunit{Ge\kern -0.1em V\!/}c}\xspace}
\newcommand{\mevcc}{\ensuremath{\aunit{Me\kern -0.1em V\!/}c^2}\xspace}
\newcommand{\gevcc}{\ensuremath{\aunit{Ge\kern -0.1em V\!/}c^2}\xspace}
\def\mum  {\ensuremath{\,\upmu\nospaceunit{m}}\xspace}

\def\fb   {\ensuremath{\aunit{fb}}\xspace}
\def\invfb   {\ensuremath{\fb^{-1}}\xspace}

\newcommand{\chisq}{\ensuremath{\chi^2}\xspace}

\newcommand{\chisqip}{\ensuremath{\chi^2_{\text{IP}}}\xspace}

\def\gsim{{~\raise.15em\hbox{$>$}\kern-.85em
          \lower.35em\hbox{$\sim$}~}\xspace}
\def\lsim{{~\raise.15em\hbox{$<$}\kern-.85em
          \lower.35em\hbox{$\sim$}~}\xspace}

\def\pt         {\ensuremath{p_{\mathrm{T}}}\xspace}

\def\ptot       {\ensuremath{p}\xspace}

\def\evtgen     {\mbox{\textsc{EvtGen}}\xspace}

\def\geant      {\mbox{\textsc{Geant4}}\xspace}

\def\photos     {\mbox{\textsc{Photos}}\xspace}

\def\pythia     {\mbox{\textsc{Pythia}}\xspace}

\def\tell1  {TELL1\xspace}
\def\ukl1   {UKL1\xspace}

\usepackage{cite} % Allows for ranges in citations
\usepackage{mciteplus}
\usepackage{longtable} % only for template; not usually to be used in PAPERs

\begin{document}

%%%%%%%%%%%%%%%%%%%%%%%%%
%%%%% Title     %%%%%%%%%
%%%%%%%%%%%%%%%%%%%%%%%%%
\renewcommand{\thefootnote}{\fnsymbol{footnote}}
\setcounter{footnote}{1}

% %%%%%%% CHOOSE TITLE PAGE--------
%\onecolumn

% ===============================================================================
% Purpose: LHCb-PAPER journal paper title page template
% Author: 
% Created on: 2010-09-25
% ===============================================================================

%%%%%%%%%%%%%%%%%%%%%%%%%
%%%%%  TITLE PAGE  %%%%%%
%%%%%%%%%%%%%%%%%%%%%%%%%
\begin{titlepage}
\pagenumbering{roman}

% Header ---------------------------------------------------
\vspace*{-1.5cm}
\centerline{\large EUROPEAN ORGANIZATION FOR NUCLEAR RESEARCH (CERN)}
\vspace*{1.5cm}
\noindent
\begin{tabular*}{\linewidth}{lc@{\extracolsep{\fill}}r@{\extracolsep{0pt}}}
\ifthenelse{\boolean{pdflatex}}% Logo format choice
{\vspace*{-1.5cm}\mbox{\!\!\!\includegraphics[width=.14\textwidth]{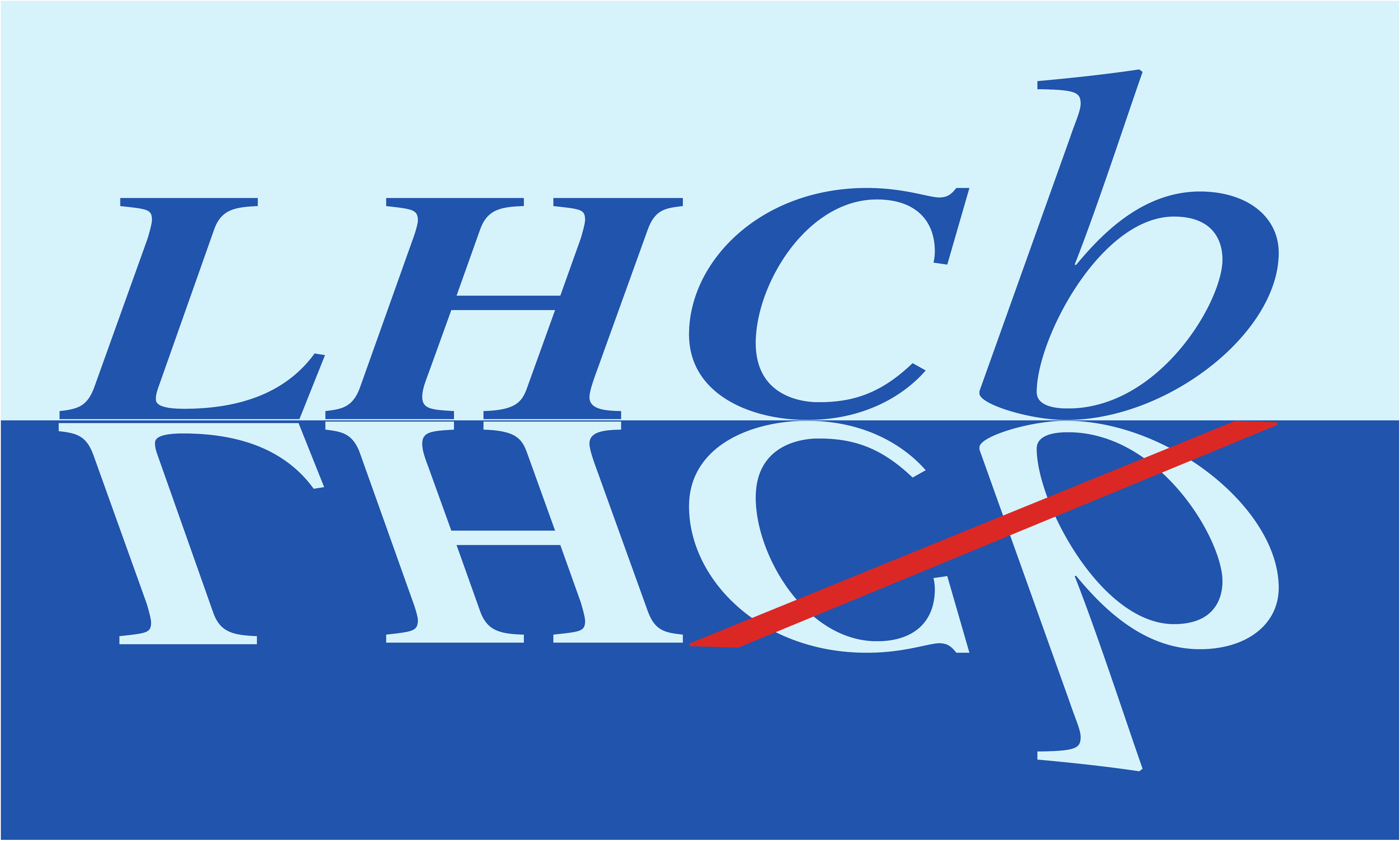}} & &}%
{\vspace*{-1.2cm}\mbox{\!\!\!\includegraphics[width=.12\textwidth]{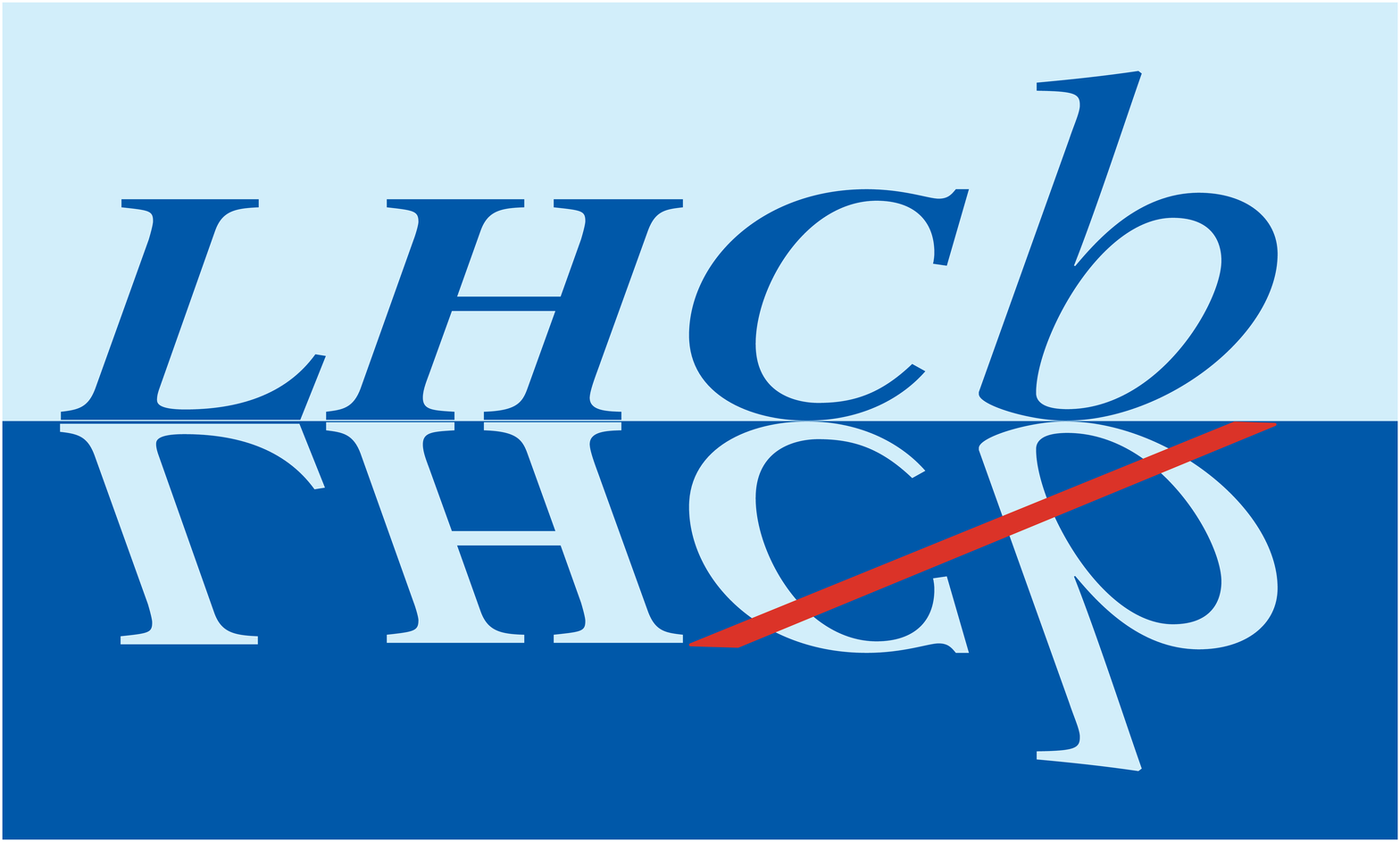}} & &}%
\\
 & & CERN-EP-2020-190 \\  % ID 
 & & LHCb-PAPER-2020-032 \\  % ID 
 & & 28 October 2020 \\ % Date - Can also hardwire e.g.: 23 March 2010
 & & \\
% not in paper \hline
\end{tabular*}

\vspace*{0.5cm}

% Title --------------------------------------------------
{\normalfont\bfseries\boldmath\huge
\begin{center}
% DO NOT EDIT HERE. Instead edit macro in main.tex to keep metadata correct
  \papertitle 
\end{center}
}

\vspace*{0.5cm}

% Authors -------------------------------------------------
\begin{center}
%In the footnote, replace 'paper' by 'Letter' in case of submission to PRL or PLB 
% Edit macro in main.tex to keep metadata correct
\paperauthors\footnote{Authors are listed at the end of this paper.}
\end{center}

\vspace{\fill}

% Abstract -----------------------------------------------
\begin{abstract}
  \noindent
  Using a proton-proton collision data sample collected by the LHCb experiment, corresponding to an integrated 
luminosity of 8.5\invfb, the observation of a new excited $\Xibz$ resonance decaying to the $\Xibm\pip$ final state is presented. The state, referred to as $\XibStarz$, has a measured mass and natural width of
\begin{align*}
m(\XibStarz) &= 6227.1^{\,+1.4}_{\,-1.5}\pm0.5\mev, \\
\Gamma(\XibStarz) &= 18.6^{\,+5.0}_{\,-4.1}\pm1.4\mev, 
\end{align*}
\noindent where the uncertainties are statistical and systematic. The production rate 
of the $\XibStarz$ state relative to that of the $\Xibm$ baryon in the kinematic region $2<\eta<5$ and $\pt<30$\gev is measured to be
\begin{align*}
\frac{f_{\XibStarz}}{f_{\Xibm}}\BF(\XibStarz\to\Xibm\pip) = 0.045\pm0.008\pm0.004,
\end{align*}
\noindent where  $\BF(\XibStarz\to\Xibm\pip)$ is the branching fraction of the decay, and $f_{\XibStarz}$ and $f_{\Xibm}$ represent fragmentation fractions. 
Improved measurements of the mass and natural width of the previously observed $\XibStarm$ state, along with the mass of the $\Xibm$ baryon, are also reported. Both measurements are significantly more precise than, and consistent with, previously reported values.
\end{abstract}

\vspace*{1.0cm}

\begin{center}
  Published in Phys. Rev. D103 (2021) 012004
\end{center}

\vspace{\fill}

{\footnotesize 
% Edit macro in main.tex to keep metadata correct
\centerline{\copyright~\papercopyright. \href{\paperlicenceurl}{\paperlicence}.}}
\vspace*{2mm}

\end{titlepage}

%%%%%%%%%%%%%%%%%%%%%%%%%%%%%%%%
%%%%%  EOD OF TITLE PAGE  %%%%%%
%%%%%%%%%%%%%%%%%%%%%%%%%%%%%%%%

%  empty page follows the title page ----
\newpage
\setcounter{page}{2}
\mbox{~}

%\twocolumn
% %%%%%%%%%%%%% ---------
\newcommand\TTAB{\rule{0pt}{2.6ex}}       % Top strut
\newcommand\BTAB{\rule[-1.2ex]{0pt}{0pt}} % Bottom strut
\newcommand\TOPTAB{\rule[0.5ex]{0pt}{1.0ex}} % Bottom strut

\renewcommand{\thefootnote}{\arabic{footnote}}
\setcounter{footnote}{0}

%%%%%%%%%%%%%%%%%%%%%%%%%%%%%%%%
%%%%%  Table of Content   %%%%%%
%%%%%%%%%%%%%%%%%%%%%%%%%%%%%%%%
%%%% Uncomment if desired
%\tableofcontents
\cleardoublepage

%%%%%%%%%%%%%%%%%%%%%%%%%
%%%%% Main text %%%%%%%%%
%%%%%%%%%%%%%%%%%%%%%%%%%

\pagestyle{plain} % restore page numbers for the main text
\setcounter{page}{1}
\pagenumbering{arabic}

%% Uncomment during review phase. 
%% Comment before a final submission.
%%\linenumbers

\section{Introduction}

In the constituent quark model~\cite{GellMann:1964nj,Zweig:352337}, baryonic states
form multiplets according to the symmetry of their flavor, spin and spatial wave functions.
The masses, natural widths and decay modes of these states give insight into their internal structure~\cite{Klempt:2009pi}. 
The $\Xibz$ and $\Xibm$ states form an isodoublet of $bsq$ bound states, where
$q$ is a $u$ or $d$ quark, respectively. Three such isodoublets, which are neither radially nor orbitally excited, should exist~\cite{Ebert:1995fp}, 
and include the $\Xib$ state with spin $j_{qs}=0$ and $J^P=(1/2)^+$, the $\Xib^{\prime}$ with $j_{qs}=1$ and $J^P=(1/2)^+$,
and the $\Xib^{*}$ with $j_{qs}=1$ and $J^P=(3/2)^+$. Here, $j_{qs}$ is the spin of the light diquark system 
$qs$, and $J^P$ represents the spin and parity of the state. Three of the four $j_{qs}=1$ states
have been observed through their decays to $\Xibz\pim$ and $\Xibm\pip$ final states~\cite{LHCb-PAPER-2014-061,Chatrchyan:2012ni,LHCb-PAPER-2016-010}. 

Beyond these lowest-lying $\Xib$ states, a spectrum of heavier states is 
expected~\cite{Ebert:2011kk,Ebert:2007nw,Roberts:2007ni,Garcilazo:2007eh,Chen:2014nyo,Mao:2015gya,PhysRevD.87.034032,Karliner:2008sv,Wang:2010it,Valcarce:2008dr,Vijande:2012mk,Wang:2017kfr,Wang:2017goq,Chen:2016phw,Thakkar:2016dna}, where there are either radial or orbital excitations amongst the constituent quarks.
Recently, peaks in the $\Lb\Km$ and $\Xibz\pim$ invariant-mass spectra corresponding to a mass of 6227\mev\footnote{Natural units with $c=1$ are used throughout this paper.} have been reported~\cite{LHCb-PAPER-2018-013},
and subsequent constituent quark model~\cite{Wang:2018fjm,Chen:2018orb,Chen:2018vuc,Aliev:2018lcs,Cui:2019dzj,Yang:2019cvw,Yang:2020zrh} and 
quark-diquark~\cite{Faustov:2020gun,Faustov:2018vgl,Jia:2019bkr,Kim:2020imk} analyses show that this state is consistent with a $P$-wave $\Xibm$ 
excitation. Alternative investigations argue that the state could also be wholly or partially molecular in nature~\cite{Zhu:2020lza,Yu:2018yxl,Nieves:2019jhp,Huang:2018bed}.
More information on the observed states, or observation of additional excited 
beauty-baryon states, will provide additional input for these theoretical investigations.

In this article, the observation of a new beauty-baryon 
resonance, referred to as $\XibStarz$, is reported using samples of proton-proton ($pp$) collision data collected with the \lhcb experiment
at center-of-mass energies of $\sqrt{s}=7$, 8\tev (Run 1) and 13\tev (Run 2), corresponding to integrated luminosities of 1.0, 2.0 and 5.5\invfb, respectively.
The resonance is seen through its decay to the $\Xibm\pip$ final state, where the $\Xibm$ baryon is reconstructed in the
fully hadronic decay channels $\Xicz\pim$ and $\Xicz\pim\pip\pim$, with \mbox{$\Xicz\to p\Km\Km\pip$}. 
Charge-conjugate processes are implicitly included throughout this paper. 

Using the 13\tev data, the production rate of the $\XibStarz$ state is measured relative to that of the $\Xibm$ baryon as
\begin{align}
R(\Xibm\pip)\equiv \frac{f_{\XibStarz}}{f_{\Xibm}}\BF(\XibStarz\to\Xibm\pip).
\end{align}
\noindent Here, $f_{\XibStarz}$ and $f_{\Xibm}$ are the fragmentation fractions for $b\to\XibStarz$ and $b\to\Xibm$, which include contributions from the decays of higher-mass $b$-hadrons, and \mbox{$\BF(\XibStarz\to\Xibm\pip)$} is the branching fraction of the decay. 

The same $pp$ collision data set is used to improve the precision on the mass and width of the
recently observed $\XibStarm$ state~\cite{LHCb-PAPER-2018-013} using 
the $\XibStarm\to\Lb\Km$ decay mode. The analysis presented here benefits greatly from the larger data sample,
but also by using both $\Lb\to\Lc\pim$ and $\Lb\to\Lc\pim\pip\pim$ decays, leading to about a four-fold increase in the $\Lb$ yield over that which was used in Ref.~\cite{LHCb-PAPER-2018-013}. 

Lastly, with the large samples of $\Xibm$ and $\Lb$ decays obtained in this analysis, 
the most precise measurement of the $\Xibm$ mass to date is presented.  The $\Xibm$ mass obtained in this
analysis is then used to obtain the mass of the $\XibStarz$ resonance.

\section{Detector and simulation}

The \lhcb detector~\cite{LHCb-DP-2008-001,LHCb-DP-2014-002} is a single-arm forward
spectrometer covering the \mbox{pseudorapidity} range $2<\eta <5$,
designed for the study of particles containing \bquark or \cquark
quarks. The detector includes a high-precision tracking system
consisting of a silicon-strip vertex detector surrounding the $pp$
interaction region, a large-area silicon-strip detector located
upstream of a dipole magnet with a bending power of about
$4{\mathrm{\,Tm}}$, and three stations of silicon-strip detectors and straw
drift tubes placed downstream of the magnet.
The tracking system provides a measurement of the momentum, \ptot, of charged particles with
a relative uncertainty that varies from 0.5\% at low momentum to 1.0\% at 200\gev.
The minimum distance of a track to a primary vertex (PV), the impact parameter (IP), 
is measured with a resolution of $(15+29/\pt)\mum$,
where \pt is the component of the momentum transverse to the beam, in\,\gev.
Different types of charged hadrons are distinguished using information
from two ring-imaging Cherenkov detectors.
Photons, electrons and hadrons are identified by a calorimeter system consisting of
scintillating-pad and preshower detectors, an electromagnetic
and a hadronic calorimeter. Muons are identified by a
system composed of alternating layers of iron and multiwire
proportional chambers.
The online event selection is performed by a trigger
which consists of a hardware stage, based on information from the calorimeter and muon
systems, followed by a software stage, which applies a full event reconstruction.

Simulation is required to model the effects of the detector acceptance and the
imposed selection requirements. It is also used to determine the expected invariant-mass resolution.
In the simulation, $pp$ collisions are generated using \pythia~\cite{Sjostrand:2007gs,*Sjostrand:2006za} 
with a specific \lhcb configuration~\cite{LHCb-PROC-2010-056}.
Decays of unstable particles are described by \evtgen~\cite{Lange:2001uf}, in which final-state
radiation is generated using \photos~\cite{Golonka:2005pn}.
The interaction of the generated particles with the detector, and its response,
are implemented using the \geant toolkit~\cite{Allison:2006ve, *Agostinelli:2002hh} as described in
Ref.~\cite{LHCb-PROC-2011-006}. 

To improve the agreement of the simulation with the data in modeling the kinematics of beauty baryons within the acceptance of the LHCb detector,
the simulated beauty-baryon momentum components, $\pt$ and $p_z$, are transformed to match the distributions obtained
from background-subtracted data~\cite{Pivk:2004ty}. Here, $p_z$ is the momentum component along the beam axis.
In particular, the $\pt$ and $p_z$ are transformed according to
\begin{align}
\pt\to\pt^{\prime}&=\exp(\kappa_{\rm T}\log(\pt)),  \nonumber \\
p_z\to p_z^{\prime}&=\exp(\kappa_z\log(p_z)).
\label{eq:ptpz_map}
\end{align}
\noindent For the $\Lb$ and $\Xibm$ simulations, the values $\kappa_{\rm T}=0.98$ and $\kappa_z=0.99$ bring the simulated $\pt$ and $p_z$ 
distributions into good agreement with those of the data, while for the $\XibStarz$ and $\XibStarm$ simulations, the values 
$\kappa_T=0.99$ and $\kappa_z=1.0$ are found. Values of $\kappa$ less than
unity indicate that the given momentum component needs to be scaled to lower values to bring the simulation into agreement with the data.
In the optimization of specific selections and the determination of selection efficiencies, these tunings are employed, 
as discussed below.

The particle identification (PID) response of charged hadrons produced in simulated signal decays is obtained from dedicated calibration samples from the data where no PID requirements are imposed~\cite{LHCb-DP-2012-003,LHCb-DP-2018-001}.
The $\Dstarp\to\Dz\pip$ mode is used for the $\Km$ and $\pip$ meson PID responses and the $\Lb\to\Lc\pim$ and $\Lz\to p\pim$ decays
are used for the proton PID response. Each final-state signal hadron has its PID response drawn from a three-dimensional probability
distribution function that depends on the hadron's $\ptot$ and $\pt$, and the number of reconstructed charged particles in the event. 

\section{Selection requirements}
\subsection{\boldmath{\Xibm} and $\Lb$ baryon selections}
The $\Xibm$ candidates are reconstructed using the $\Xicz\pim$ and $\Xicz\pim\pip\pim$ decay modes,
while the $\Lb$ sample uses the $\Lc\pim$ and $\Lc\pim\pip\pim$ final states. The charm
baryons are detected through the decays $\Xicz\to p\Km\Km\pip$ and $\Lc\to p\Km\pip$. In what follows,
$H_b$ refers to either the $\Lb$ or $\Xibm$ baryon, and $H_c$ signifies the corresponding charm baryon, $\Lc$ or $\Xicz$, 
according to the above decay sequences.

Charged hadrons used to reconstruct the $H_b$ candidates are required to
be significantly detached from all PVs in the event using the quantity $\chisqip$, which is the difference in 
$\chisq$ of the vertex fit of a given PV when the particle is included or excluded from the fit. 
Each track is required to have $\chisqip>4$, which corresponds to an IP that is at least
twice as large as the expected IP resolution. Loose PID requirements are also imposed on 
all the $H_b$ decay products to ensure that they are consistent with the intended decay sequence.

The $H_c$ candidates are required to have a good-quality vertex fit, have significant displacement 
from all PVs in the event, and satisfy the invariant-mass requirements,
$|M(p\Km\pip)-m_{\Lc}|<18$\mev and $|M(p\Km\Km\pip)-m_{\Xicz}|<15$\mev, corresponding to about three times the mass resolution. Here, and throughout this paper, $M$ represents the invariant mass of the particle(s) indicated in
parentheses, and $m$ represents the measured mass of the indicated particle, using Ref.~\cite{PDG2020} for known particles.

One or three charged pions, with total charge $-1$, are combined with $H_c$ candidates to form the $H_b$ samples.
The fitted decay vertex is required to be consistent with a single point in space, evidenced by
having good fit quality. To suppress combinatorial background,
the $H_b$ decay vertex is required to be significantly displaced from all PVs in the event
and have small $\chisqip$ to at least one PV. The $H_b$ candidates are assigned to the PV for which
$\chisqip$ is minimum.

After these selections, clear $\Xibm$ and $\Lb$ peaks can be seen in the data. The $\Lb\to\Lc\pim$ decay mode
has an excellent signal-to-background (S/B) ratio, and no further selections are applied. For the $\Xibm\to\Xicz\pim$, 
$\Xibm\to\Xicz\pim\pip\pim$ and $\Lb\to\Lc\pim\pip\pim$ decays, a boosted decision tree (BDT) discriminant~\cite{Roe,AdaBoost,Hocker:2007ht}  
is used to further improve the S/B ratio. The set of variables used by the BDT is similar for the three modes.
Those common to all three modes include: the $\chisq$ values of the fitted $H_c$ and $H_b$ decay vertices, the angle between
the $H_b$ momentum direction and the vector pointing from the PV to the $H_b$ decay vertex, the $H_b$ and $H_c$ decay
times, and for each final-state hadron, $\ptot$, $\pt$, $\chisqip$ and a PID response variable. 
For the $H_b\to H_c\pim\pip\pim$ modes, three additional variables are included: $M(\pim\pip\pim)$,
the $\chisq$ of the $\pim\pip\pim$ vertex fit, and the $\chisq$ of the vertex separation between the $3\pi$ vertex and the 
associated PV. The BDT is trained using simulated decays for the signal distributions in these variables, and the background distributions are taken from 
a combination of the $H_c$ or $H_b$ mass sidebands in data. The requirements on the
BDT discriminant are chosen based on optimizing the product of signal efficiency and signal purity. The
resulting BDT selection requirement is $\sim$100\%, 94\% and 93\%  efficient for 
$\Xibm\to\Xicz\pim$, $\Xibm\to\Xicz\pim\pip\pim$ and $\Lb\to\Lc\pim\pip\pim$ signal decays, while
suppressing the combinatorial background by factors of about 3, 8 and 6, respectively.

In anticipation that the $\Xibm\to\Xicz\pim$ decay mode will be used to measure the relative production rate,
$R(\Xibm\pip)$, $\Xibm$ candidates are restricted to lie in the kinematic region $\pt<30$\gev and $2<\eta<5$;
this selection retains 99.7\% of the signal decays.

\begin{figure}[tb]
  \begin{center}
    \includegraphics[width=0.49\linewidth]{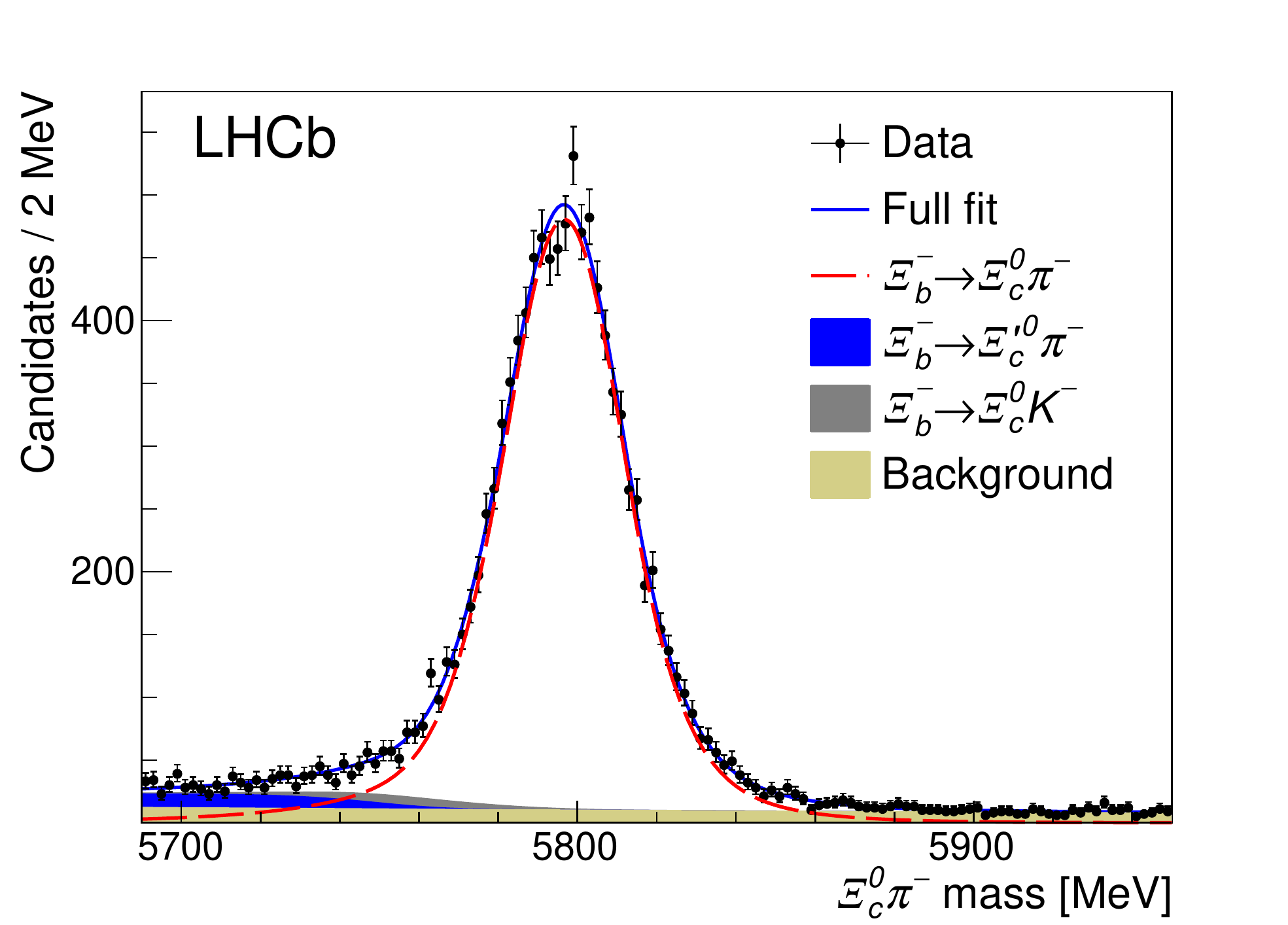}
    \includegraphics[width=0.49\linewidth]{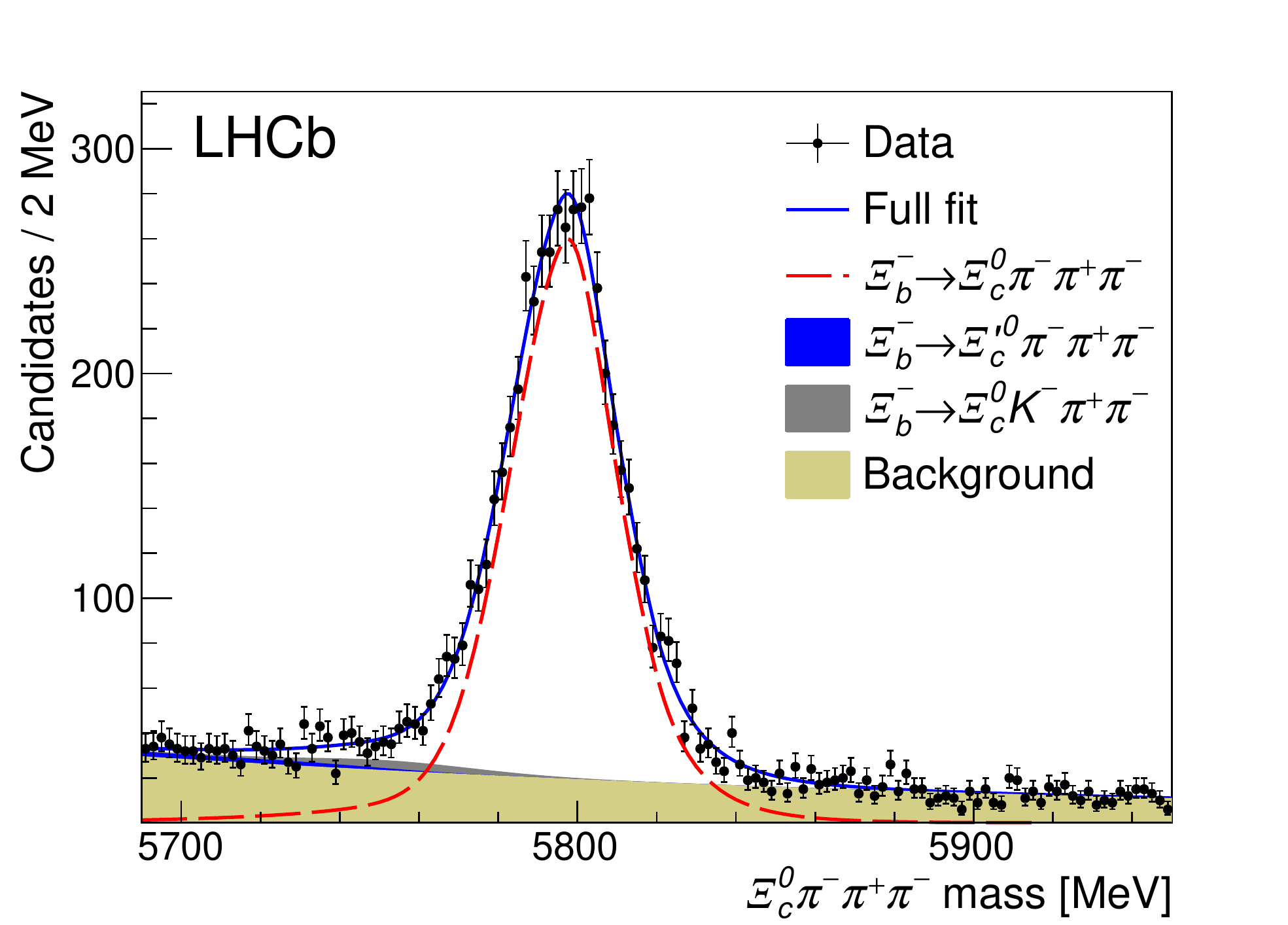}
    \vspace*{-0.5cm}
  \end{center}
  \caption{
    \small{Invariant-mass spectra for (left) $\Xibm\to\Xicz\pim$ and (right) $\Xibm\to\Xicz\pim\pip\pim$ 
candidates after all selection requirements. Projections of the fits to the data are overlaid.}}
  \label{fig:Xib}
\end{figure}

\begin{figure}[tb]
  \begin{center}
    \includegraphics[width=0.49\linewidth]{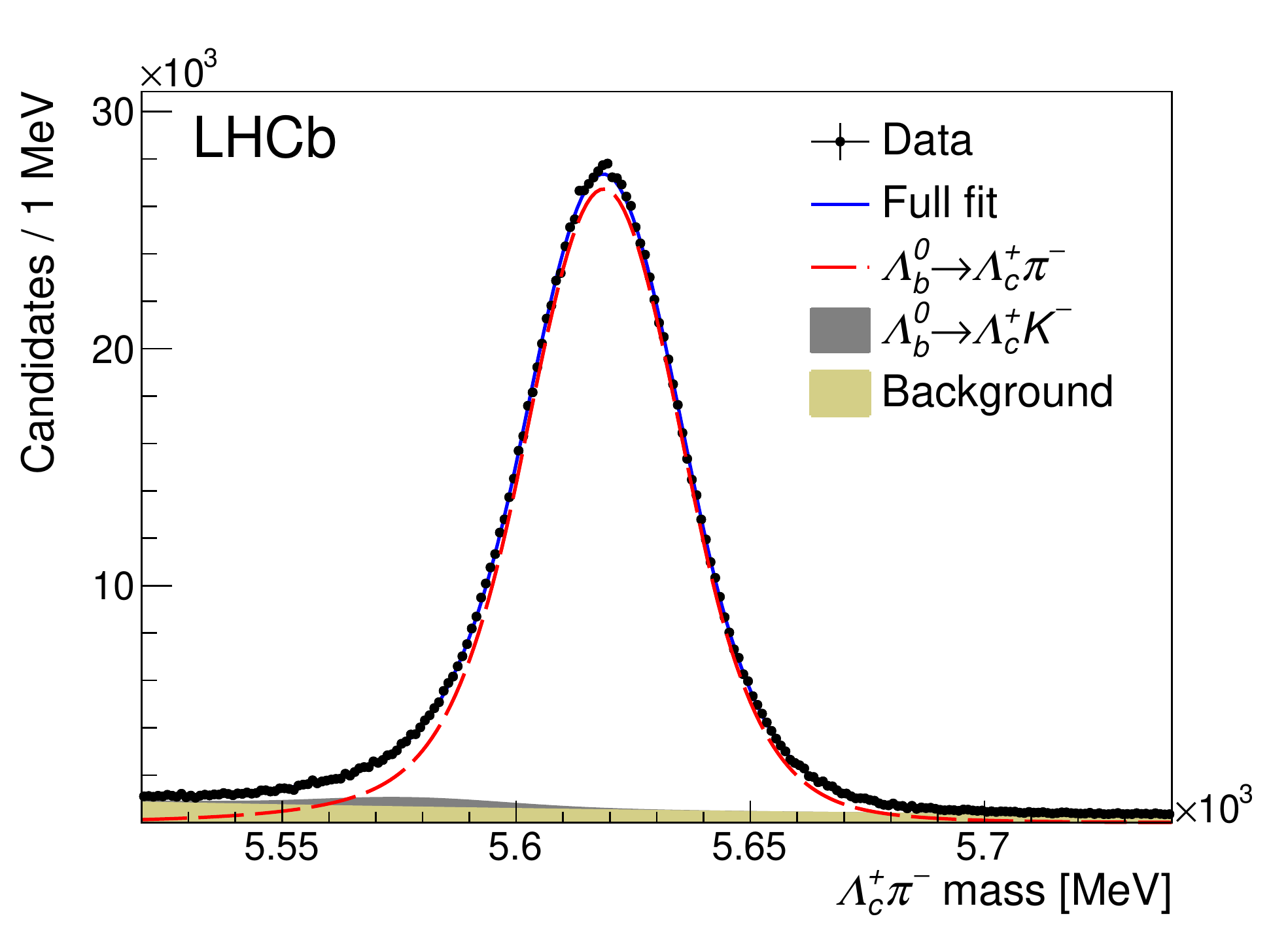}
    \includegraphics[width=0.49\linewidth]{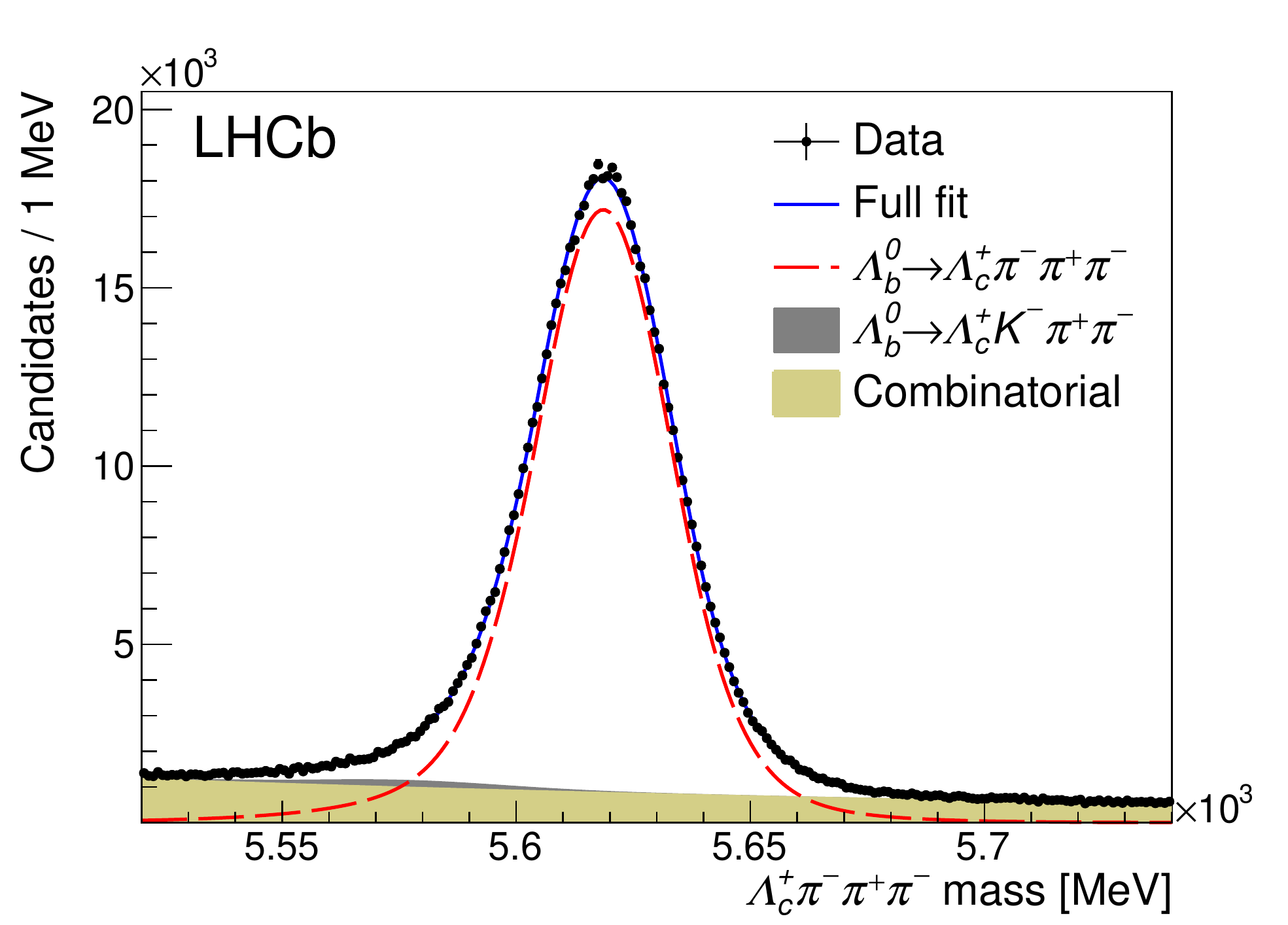}
    \vspace*{-0.5cm}
  \end{center}
  \caption{
    \small{Invariant-mass spectra for (left) $\Lb\to\Lc\pim$ and (right) $\Lb\to\Lc\pim\pip\pim$ candidates after all
selection requirements. Projections of the fits to the data are overlaid.}}
  \label{fig:Lb}
\end{figure}

With all of the selections applied, the resulting $\Xibm$ and $\Lb$ candidate invariant-mass spectra are shown in 
Figs.~\ref{fig:Xib} and~\ref{fig:Lb}, respectively. The fits, as described below, are overlaid.

\subsection{\boldmath{\XibStarz selection}}
The $\XibStarz$ candidates are formed by combining a $\Xibm$ candidate with a $\pip$ meson consistent with coming from the same PV.
The $\Xibm\to\Xicz\pim$  and $\Xibm\to\Xicz\pim\pip\pim$ candidates are required to have their masses in the intervals 
$5737<M(\Xicz\pim)<5847$\mev and 
$5750<M(\Xicz\pim\pip\pim)<5840$\mev, respectively, corresponding to about three times the mass resolution about the $\Xibm$ mass~\cite{PDG2020}.

The majority of particles from the PV are pions, and therefore only a loose requirement is applied to the pion PID hypothesis, 
sufficient to render the contribution from misidentified kaons and protons to be at the few percent level. To suppress background from
random $\pip$ mesons, which tend to have lower $\pt$ than those from $b$-hadron decays, the selection on the $\pt$ of the $\pip$ candidate
is optimized as follows. The Punzi figure-of-merit~\cite{Punzi:2003bu} FOM$\,=\epsilon(\pt)/(\sqrt{N_B(\pt)}+a/2)$ 
with $a=5$ is used, where $\epsilon(\pt)$ and $N_B(\pt)$ are the signal efficiency and background yield as a function of the applied $\pip$ 
meson $\pt$ requirement. For the signal efficiency, $\epsilon(\pt)$, the $\pip$ meson $\pt$ is scaled by the ratio $\pt^{\prime}/\pt$, as given 
in Eq.~(\ref{eq:ptpz_map}). The optimal requirements are $\pt>700$\mev and 900\mev for the $\Xibm\to\Xicz\pim$ and $\Xibm\to\Xicz\pim\pip\pim$ modes,
respectively. The higher $\pt$ requirement on the latter is due to the higher average momentum required of the $\XibStarz$ baryon
in order for all of its decay products to be within the \lhcb detector acceptance.
These selections provide an expected signal 
efficiency of about 55\% and reduce the background by an order of magnitude. 

\subsection{\boldmath{\XibStarm selection}}
The $\XibStarm$ candidates are formed by combining $\Lb$ candidates in the mass interval 5560--5670\mev 
and $\Km$ candidates consistent with emerging from the same PV.
A similar optimization to that discussed above is performed to determine the optimal $\pt$ requirement on the $\Km$ candidate.
%Here again, the $\Km$ meson $\pt$ is scaled according to Eq.~\ref{eq:ptpz_map} to better reflect the expected %$\pt$ distribution in data.
A loose PID requirement on the $\Km$ candidate is applied in advance, which suppresses about 80\% of the misidentified $\pim$ background.
Since the $\XibStarm$ state is established, the optimization uses FOM$\,=N_S(\pt)/\sqrt{N_S(\pt)+N_B(\pt)}$, where
$N_S(\pt)=\epsilon(\pt)N_{S0}$ is the expected signal yield based on an initial signal yield estimate, $N_{S0}$, and the
efficiency, $\epsilon(\pt)$, obtained from simulation.  The background yield, $N_B$, is obtained from wrong-sign 
$\Lb\Kp$ combinations. The optimal requirement is $\pt>1000$\mev. The efficiency of this selection is about 40\% and reduces
the combinatorial background by a factor of ten. 

With the $\pt>1000$\mev requirement applied, a more refined optimization is performed on the $\Km$ PID requirement.
The PID tuning for the 7 and 8\tev data differs from that of the 13\tev data~\cite{LHCb-DP-2018-001}, so different requirements are imposed.
Using the same FOM as above, except with the PID variable used in place of the $\pt$, tighter PID requirements are imposed.
The optimal PID requirement on the $\Km$ candidate provides an efficiency of
80\% (95\%) while suppressing the background by a factor of 2 (1.6) for the Run 1 (Run 2) data samples. The same $\pt$ and 
PID requirements are applied to the $\Km$ candidate in both the $\Lb\to\Lc\pim$ and $\Lb\to\Lc\pim\pip\pim$ samples. 

\section{Fits to the data}
\subsection{\boldmath{Fits to the \Xibm and \Lb samples}}
An extended binned maximum-likelihood fit is performed to determine the $\Xibm$ and $\Lb$ signal yields in the
peaks shown in Figs.~\ref{fig:Xib} and~\ref{fig:Lb}. The distributions are
described by the sum of a signal function and three (two for $\Lb$) background shapes to determine the signal yields. 
The signal shapes are described by the sum of two Crystal Ball functions~\cite{Skwarnicki:1986xj} with a common value for the
peak mass. For the $\Xibm$ modes, the signal shapes are fixed to the values obtained from simulation, except for
the widths, which are allowed to vary freely in the fit. For the $\Lb$ modes, the signal yields in data are
significantly larger than in the simulated samples, and thus all signal shape parameters are freely varied in the fit.
For both the $\Lb$ and $\Xibm$ modes, there is background from $H_b\to H_c\Km(\pip\pim)$ decays, where the kaon is misidentified as
a pion. This Cabibbo-suppressed (CS) contribution is small compared to the Cabibbo-favored (CF) $H_b\to H_c\pim\pip\pim$ decay.
The CS to CF signal yield ratio is fixed to 1.8\% based upon the PID efficiency of the $\Km$ meson to pass the $\pim$ PID requirement
and the assumption that the CS/CF ratio of branching fractions is 7.3\%, as is the case for $\BR(\Lb\to\Lc\Km)/\BR(\Lb\to\Lc\pim)$~\cite{LHCB-PAPER-2013-056}.
For the $\Xibm$ modes, there is also a background
contribution from $\Xibm\to\XicPrime\pim(\pip\pim)$ decays, where the photon from the decay $\XicPrime\to\Xicz\gamma$ is not
considered. The shapes of these background modes are taken from simulations and the yields are freely varied in the fit.
Lastly, the combinatorial background shapes are parametrized as an exponential function with freely varying shape parameters 
and yields.

The results of the fit are superimposed in Figs.~\ref{fig:Xib} and~\ref{fig:Lb}, and the fitted  signal yields are shown in 
Tables~\ref{tab:YieldsXib} and~\ref{tab:YieldsLb}. In total, about 1.9 million $\Lb$ and 16\,000 $\Xibm$ signal decays
are observed, with sizable contributions from 
final states containing three pions.
The number of $\Lb$ decays here is about four 
times larger than the sample used for the first measurement of the $\XibStarm$ mass and natural width~\cite{LHCb-PAPER-2018-013}.

\begin{table*}[tb]
\begin{center}
\caption{\small{Signal yields of $\Xibm$ and $\XibStarz$ decays for the full data set after all selection requirements,
and the corresponding Run 2 signal yields used for the measurement of $R(\Xibm\pip)$ at 13\tev.}}
\begin{tabular}{lccc}
\hline\hline
                               & \multicolumn{2}{c}{All data} & $\sqrt{s}=13\tev$ \\
\\[-0.18in]
$\Xibm\to$                     & $\Xicz\pim$                 &     $\Xicz\pim\pip\pim$ &  $\Xicz\pim$  \\
\\[-0.18in]
\hline
\\[-0.18in]
$N(\Xibm)$                     &  $10\,800\pm400$              & $5100\pm300$   &  $8300\pm300$ \\
\\[-0.18in]
$N(\XibStarz\to\Xibm\pip)$      &  $~~176^{\,+33}_{\,-30}$           & $~86^{\,+19}_{\,-17}$  &  $150\pm27$ \\
\\[-0.18in]
\hline\hline
\end{tabular}
\label{tab:YieldsXib}
\end{center}
\end{table*}

\begin{table*}[tb]
\begin{center}
\caption{\small{Signal yields of $\Lb$ and $\XibStarm$ decays for the full data set after all selection requirements.}}
\begin{tabular}{lcc}
\hline\hline
\\[-0.18in]
$\Lb\to$                      & $\Lc\pim$             &     $\Lc\pim\pip\pim$ \\
\\[-0.18in]
\hline
\\[-0.18in]
$N(\Lb)~[10^3]$               &  $1214\pm2$         &      $697\pm1$ \\
\\[-0.18in]
$N(\XibStarm\to\Lb\Km)$       &  $~~~1100\pm108$         &     $~\,1024\pm106$  \\
\\[-0.18in]
\hline\hline
\end{tabular}
\label{tab:YieldsLb}
\end{center}
\end{table*}

\subsection{\boldmath{Fit to the \XibStarz\to\,\Xibm\pip sample}}

\begin{figure}[tb] 
\centering
\includegraphics[width=0.49\textwidth]{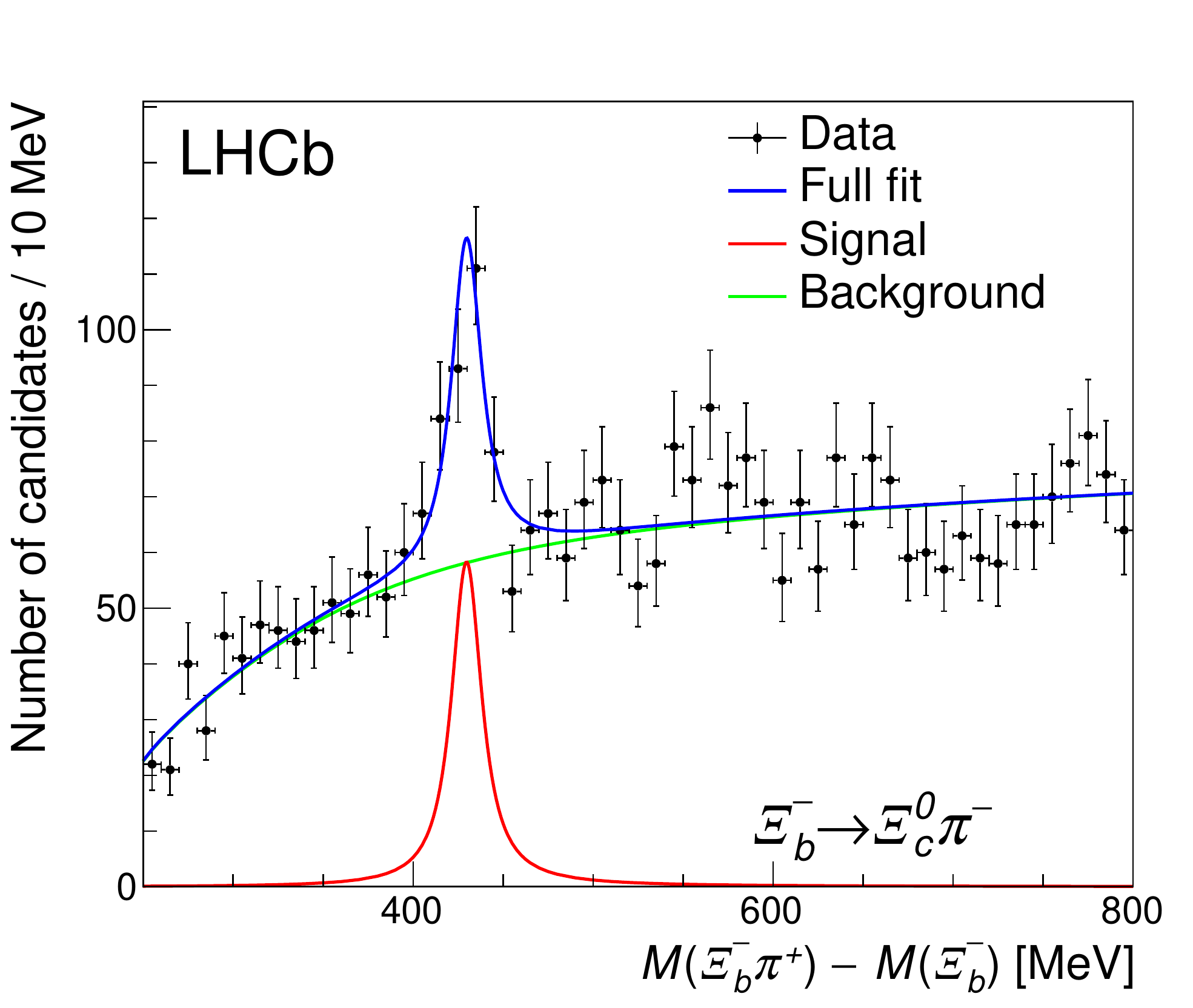}
\includegraphics[width=0.49\textwidth]{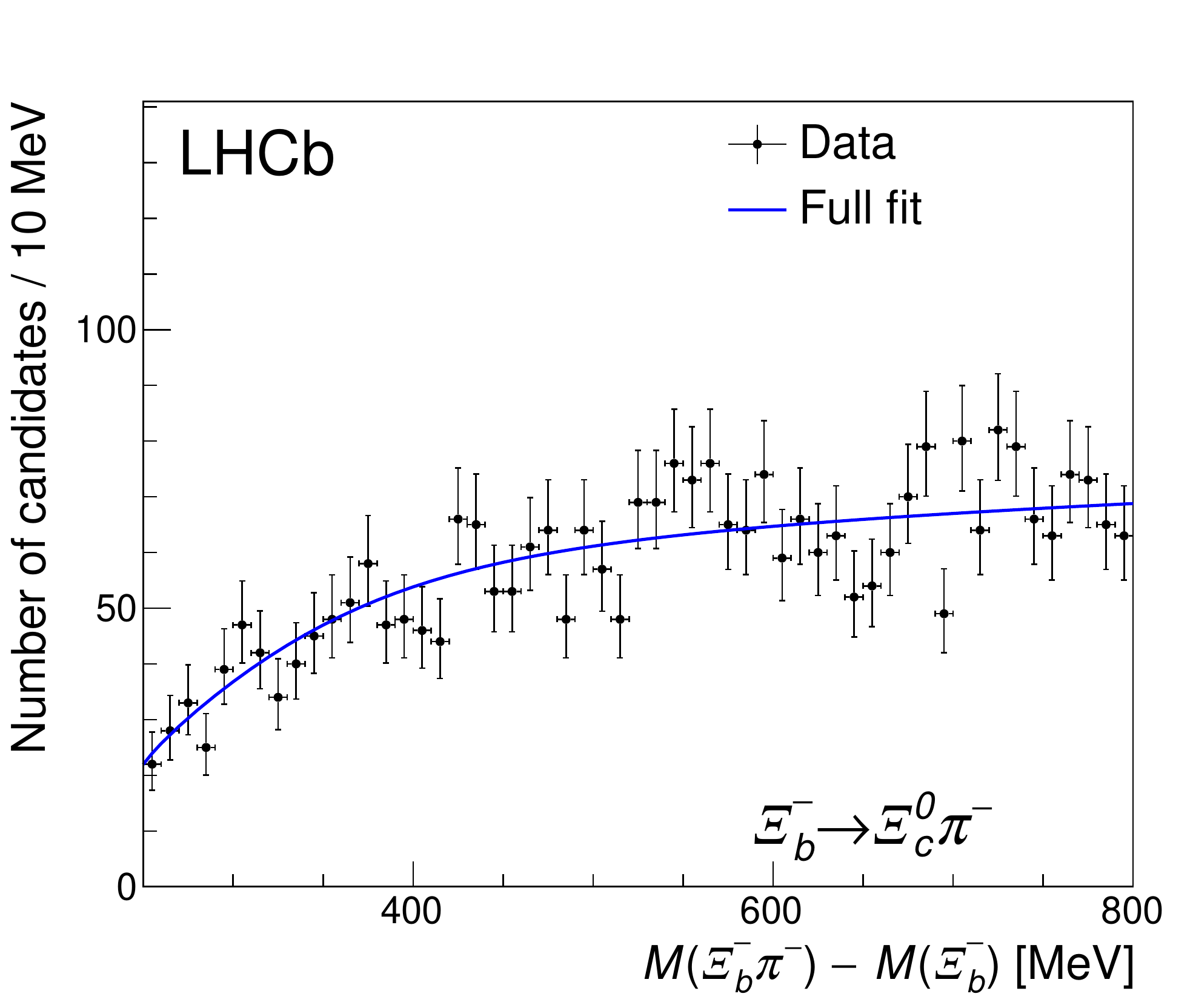}
\includegraphics[width=0.49\textwidth]{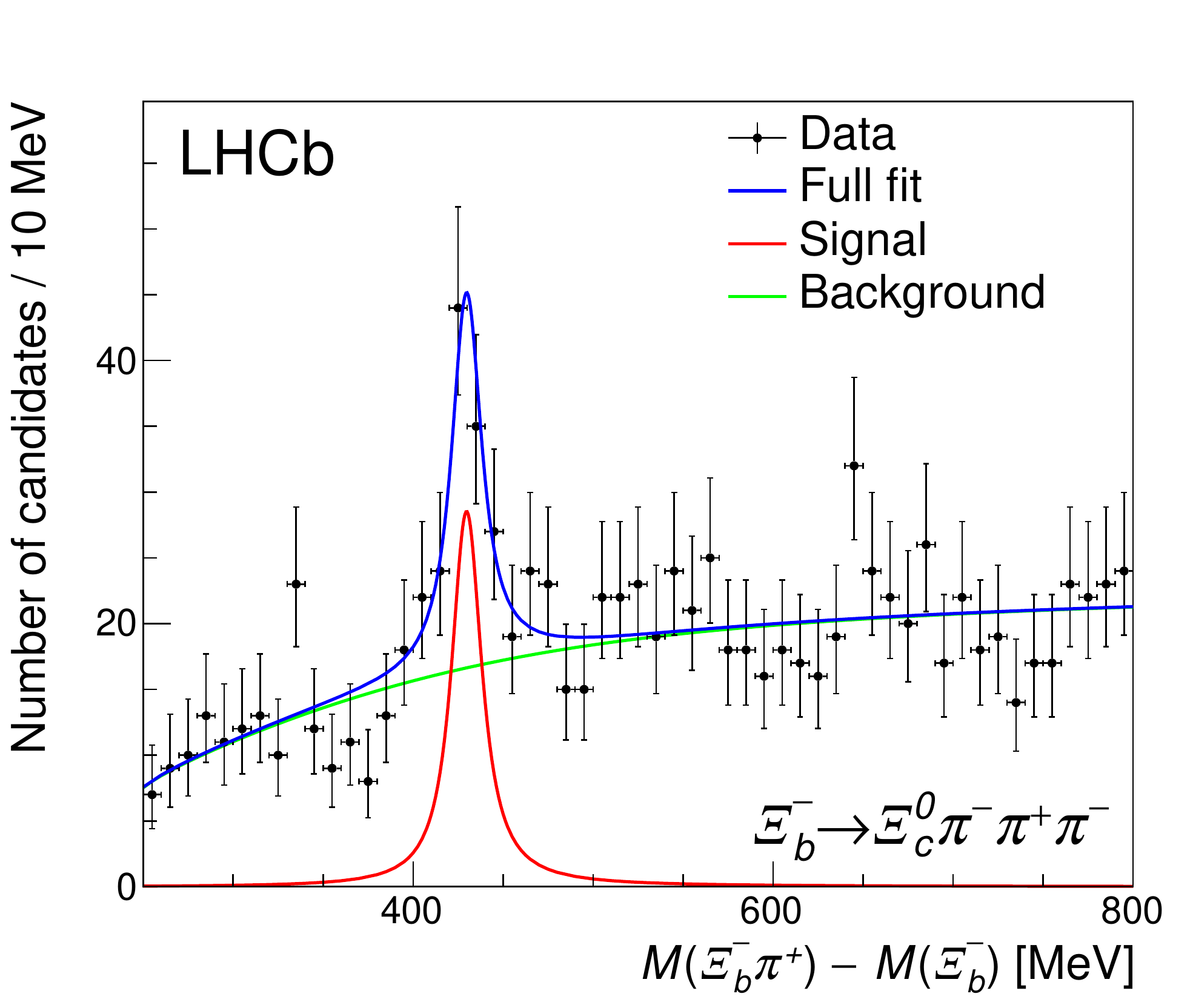}
\includegraphics[width=0.49\textwidth]{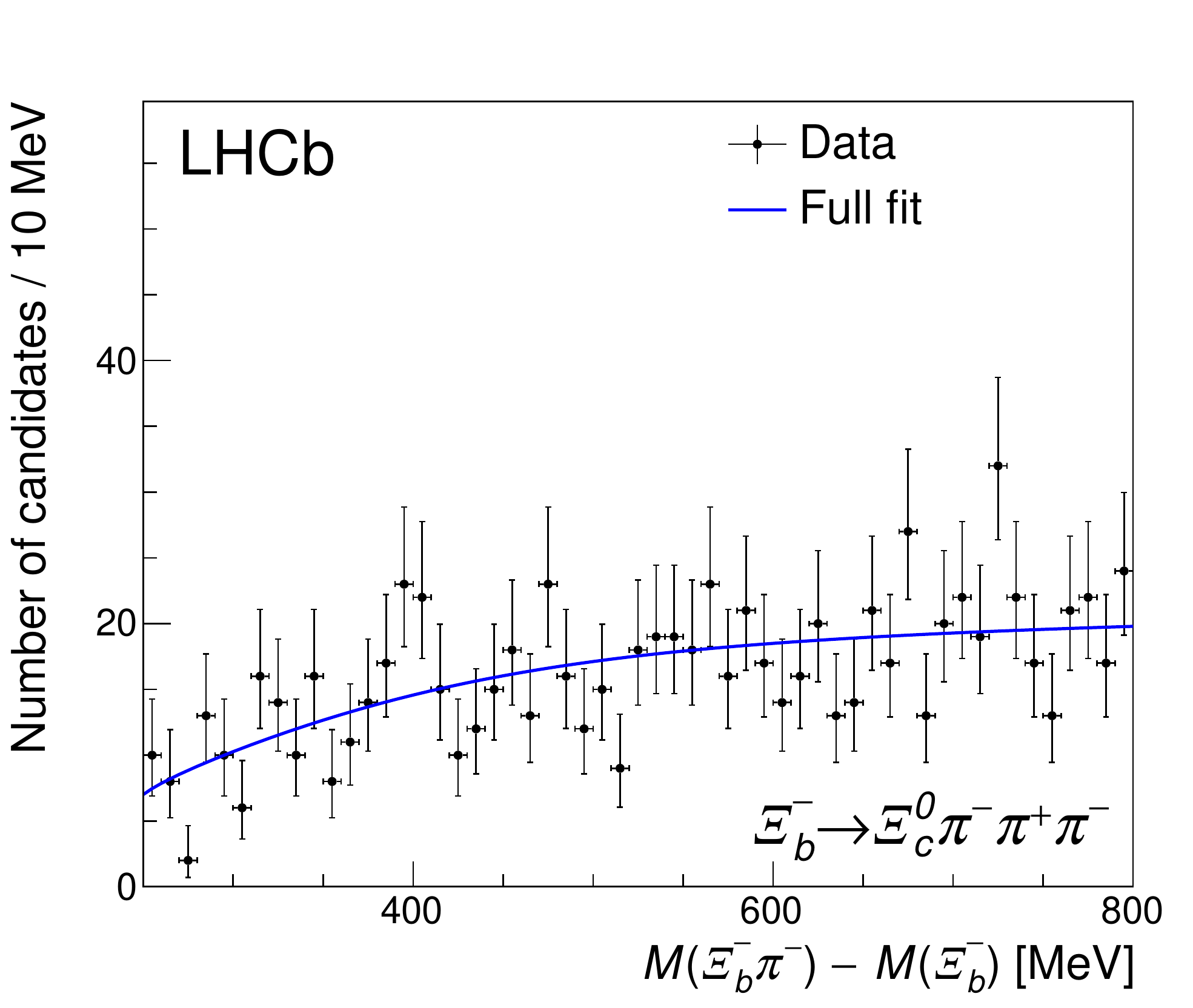}
\caption{\small{Distribution of reconstructed $\delta M_{\pi}=M(\Xibm\pip)-M(\Xibm)$ in $\XibStarz\to\Xibm\pip$ candidate decays, with
(top) $\Xibm\to\Xicz\pim$ decays, and (bottom) $\Xibm\to\Xicz\pim\pip\pim$ decays. The left column shows the
right-sign candidates and the right column shows the wrong-sign candidates. The fit projections are overlaid.}}
\label{fig:XibStarZero}
\end{figure}

To search for the $\XibStarz$ state, the mass difference, $\delta M_{\pi}=M(\Xibm\pip)-M(\Xibm)$, is used, since the
mass resolution on this difference is about eight times better than that of $M(\Xibm\pip)$. Moreover, systematic
uncertainties, particularly that due to the momentum scale calibration, are greatly reduced. 
The resulting mass difference spectra, $\delta M_{\pi}$, for both the right-sign and wrong-sign ($\Xibm\pim$) combinations are
shown in Fig.~\ref{fig:XibStarZero}. The top row shows the spectra using $\Xibm\to\Xicz\pim$ candidates and the bottom 
row shows the spectra using $\Xibm\to\Xicz\pim\pip\pim$ candidates. A clear signal is observed at the same invariant mass in both 
right-sign final states, while there are no significant structures in the wrong-sign spectra.

The $\XibStarz$ mass and natural width are obtained from a simultaneous unbinned maximum-likelihood
fit to the four $\delta M_{\pi}$ spectra. 
The signal shape is described by a $P$-wave relativistic 
Breit--Wigner function~\cite{Jackson:1964zd} with a Blatt--Weisskopf 
barrier factor~\cite{Blatt} of 3\gev$^{-1}$, convolved with a resolution function. 
The mass resolution is parametrized as the sum of two Gaussian functions with
a common mean of zero and widths that are fixed to the values obtained from simulation. The weighted average mass resolution is about 2.0\mev, which is negligible compared to the apparent width of the observed peak.
The background shape is described by a smooth threshold function with shape parameters that are common
between the right-sign and wrong-sign spectra, but independent for the $\Xicz\pim$ and $\Xicz\pim\pip\pim$ final states.
The threshold function takes the form
\begin{align}
\left(1+\tanh\left(\frac{\delta M_{\pi}-\delta M_0}{C}\right)\right)\times(\delta M_{\pi}-\delta M_0)^A.
\label{eq:background_func}
\end{align}
\noindent The parameter $\delta M_0$ represents a threshold. Due to the low signal yield, the fit
does not always converge when $\delta M_0$ is left to freely vary. 
Therefore, $\delta M_0$ is fixed to 240\mev (10\mev below the minimum of the fit range), and the value is varied as a source of systematic uncertainty. 
The parameters $A$ and $C$ are freely varied in the fit.

The projection of the fit is superimposed on the data in Fig.~\ref{fig:XibStarZero}. Using the difference in log-likelihoods between the nominal fit and a fit where the signal yield is fixed to zero, a
statistical significance of about 10$\,\sigma$ is obtained.
The $\XibStarz$ peak parameters are
\begin{align*}
\delta m_{\pi}^{\rm peak} &= 429.8^{\,+1.4}_{\,-1.5}\mev, \\
m(\XibStarz) &= 6227.1^{\,+1.4}_{\,-1.5}\mev, \\
\Gamma(\XibStarz) &= 18.6^{\,+5.0}_{\,-4.1}\mev,
\end{align*}
\noindent where the uncertainties are statistical only. The $\delta m_{\pi}^{\rm peak}$ values obtained from
independent fits to the two samples are consistent with one another, therefore justifying the combined fit. 
The $\XibStarz$ mass is obtained from
$m(\XibStarz)=\delta m_{\pi}^{\rm peak}+m(\Xibm)$, where the value $m(\Xibm)=5797.33\pm0.24$\mev obtained 
in this analysis is used, as discussed later. The fitted signal yields are shown in Table~\ref{tab:YieldsXib}.

\subsection{\boldmath{Production ratio $R(\Xibm\pip)$}}
The relative production rate is obtained from
\begin{align}
R(\Xibm\pip) = \frac{N(\XibStarz)}{N(\Xibm)~\epsilon_{\rm rel}},
\end{align}
\noindent where $N(\XibStarz)$ and $N(\Xibm)$ are the signal yields and $\epsilon_{\rm rel}$ is the relative efficiency between 
the $\XibStarz$ and $\Xibm$ selections. As the $\Xibm$ selection is common to both samples, the relative efficiency is
predominantly due to the efficiency of reconstructing and selecting the $\pip$ meson.

About 80\% of the signal is from the 13\tev data set, and therefore $R(\Xibm\pip)$ is measured using
only that subset of the data. In addition, the acceptance requirement $\pt<30$\gev and $2<\eta<5$ is applied to the reconstructed
$\XibStarz$ candidates. To obtain $N(\XibStarz)$ and $N(\Xibm)$, an alternative fit with only the 13\tev data is performed, 
with the resulting $\XibStarz$ and $\Xibm$ signal yields shown in Table~\ref{tab:YieldsXib}. The $\Xibm$ signal yield is 
obtained by integrating the $\Xibm\to\Xicz\pim$, $\Xicz\Km$, and $\XicPrime\pim$ signal shapes over the same mass interval
($5737<M(\Xicz\pim)<5847$\mev) that is used in the $\XibStarz$ selection. The $\Xicz\Km$, and $\XicPrime\pim$ components are included in the
$\Xibm$ yield because simulation shows that these misidentified $\Xibm$ decays also produce a narrow structure in the
$\delta M_{\pi}$ spectrum with approximately the same resolution as the $\Xicz\pim$ signal.

The relative signal efficiency is obtained from the tuned simulation, from which the value ${\epsilon_{\rm rel}=(40.0\pm0.5)\%}$ is obtained, where the uncertainty is due to the finite
simulated sample sizes. Much of the efficiency loss is due to the $\pt>700\mev$ requirement; with a less stringent requirement of $\pt>200\mev$, the 
relative efficiency is 75\%. The efficiency includes a correction factor of $0.978\pm0.021$, which accounts for a slightly lower tracking
efficiency in data than in simulation, as determined from an inclusive $\jpsi\to\mup\mun$ calibration sample~\cite{LHCb-DP-2013-002}, weighted to match the kinematics of the $\pip$ meson from the $\XibStarz$ decay.

With the signal yields in Table~\ref{tab:YieldsXib} and the above value of $\epsilon_{\rm rel}$, it is found that
\begin{align*}
R(\Xibm\pip) = 0.045\pm0.008,
\end{align*}
\noindent where the uncertainty is statistical only. 

\subsection{\boldmath{Fit to the \XibStarm\to\,\Lb\Km sample}}

\begin{figure}[tb] 
\centering
\includegraphics[width=0.49\textwidth]{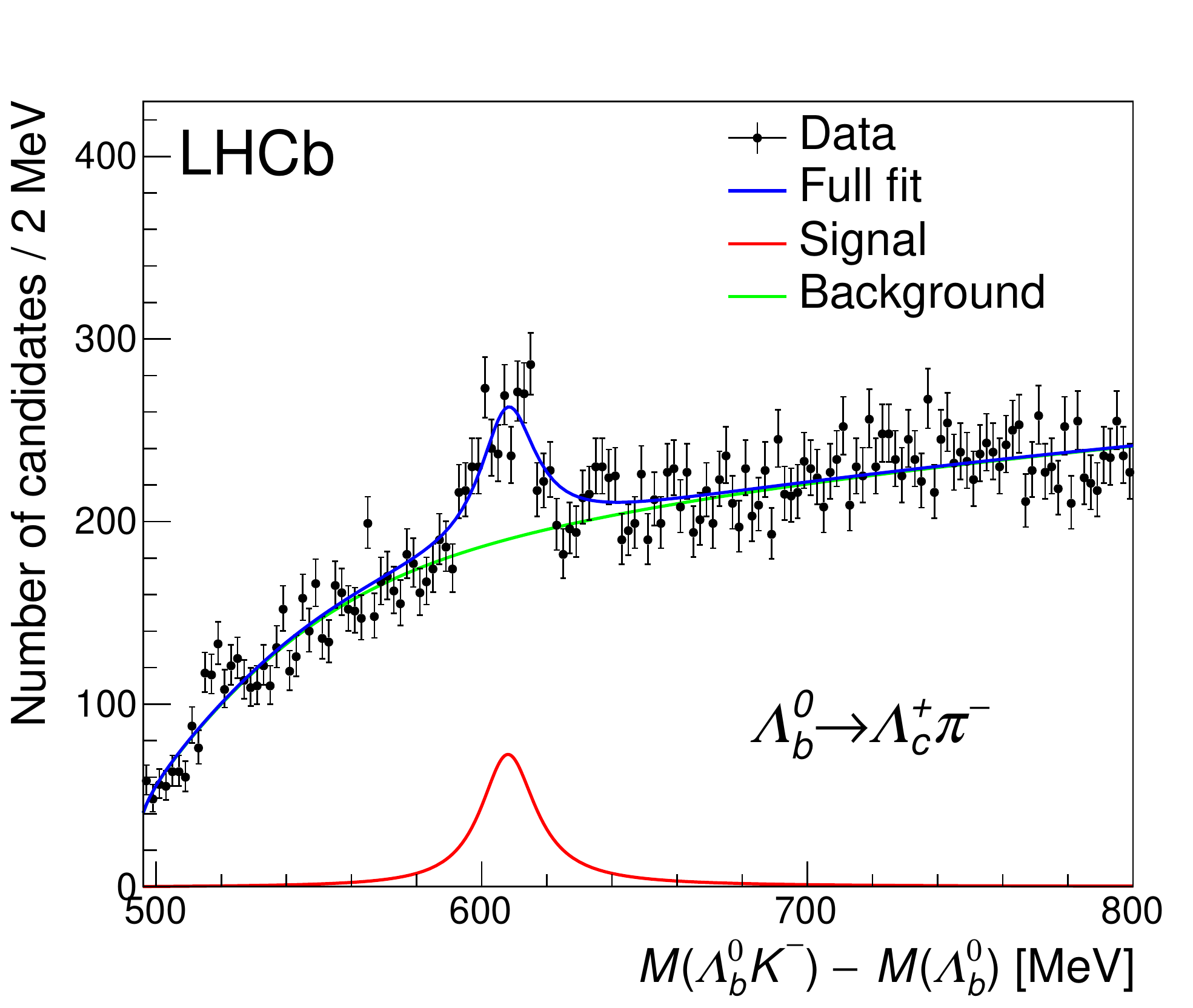}
\includegraphics[width=0.49\textwidth]{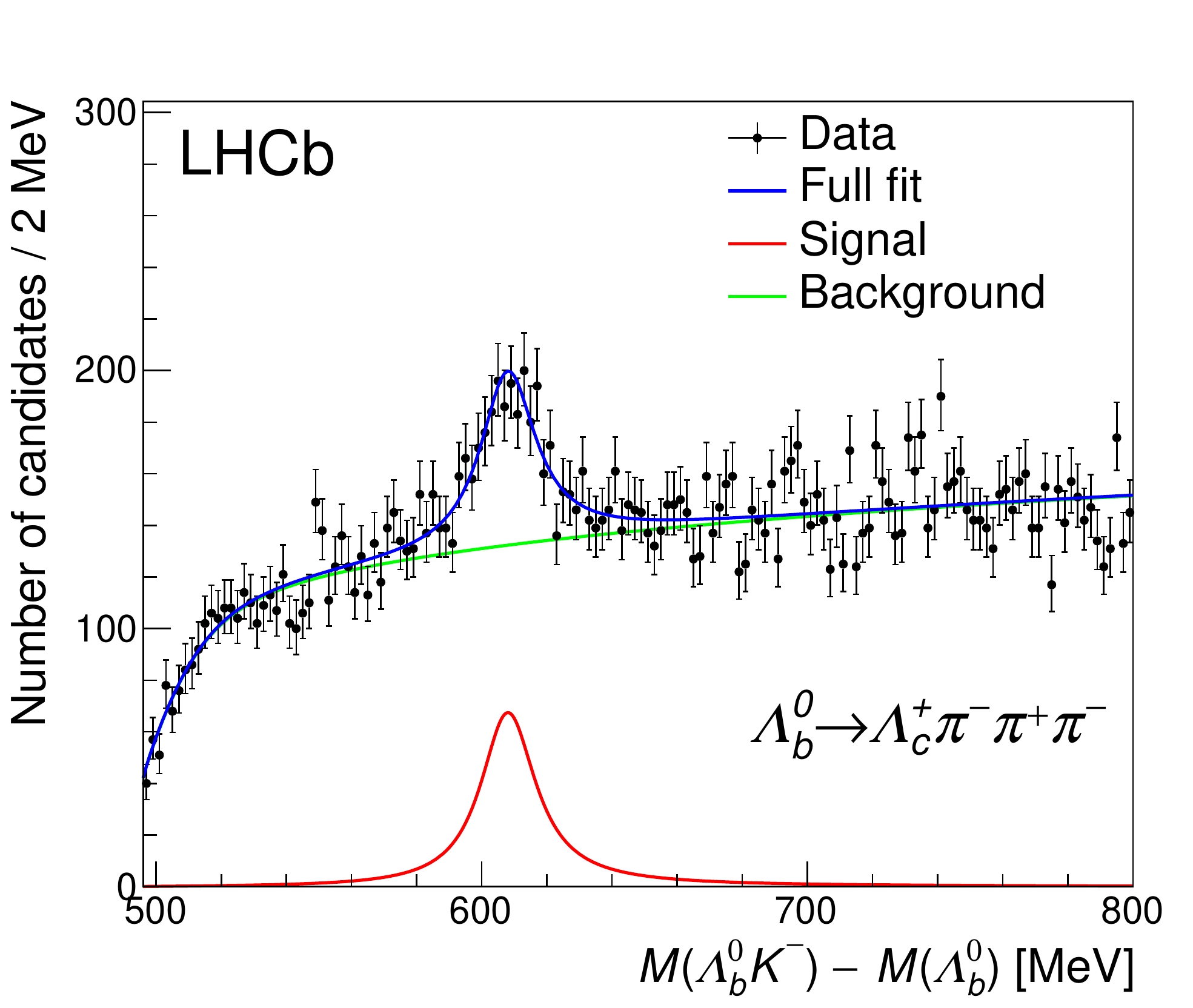}
\caption{\small{Distribution of reconstructed $\delta M_K = M(\Lb\Km)-M(\Lb)$ in $\XibStarm\to\Lb\Km$ candidate decays, with
(left) $\Lb\to\Lc\pim$ and (right) $\Lb\to\Lc\pim\pip\pim$ candidates.
The fit projections are overlaid.}}
\label{fig:XibStar2LbK_2MeV}
\end{figure}
The spectra of mass differences, $\delta M_K=M(\Lb\Km)-M(\Lb)$, are shown in Fig.~\ref{fig:XibStar2LbK_2MeV} for the $\Lb\to\Lc\pim$ and $\Lb\to\Lc\pim\pip\pim$ modes. As with the $\XibStarz$ signal fit, an unbinned 
extended maximum-likelihood fit is performed. The wrong-sign spectra are not considered in the fit, since the 
$\delta M_K$ background shape for the wrong-sign is visibly different from that of the right-sign. 
As for the $\XibStarz$ fit, the signal shape is described by
a $P$-wave relativistic Breit--Wigner function with a Blatt--Weisskopf 
barrier factor convolved with a resolution function. The mass resolution
is described by the sum of two Gaussian functions with 
a common mean of zero and widths that are fixed to the values obtained from simulation. The weighted-average width is about 1.4\mev, which is small compared to the expected natural width of
the signal peak. The background shape is given by the same functional form as Eq.~(\ref{eq:background_func}), with the replacement
$\delta M_{\pi}\to\delta M_K$ and $\delta M_0$ is fixed to the kaon mass~\cite{PDG2020}; the parameters $A$ and $C$ are freely varied in the fit.

The fit projections are superimposed to the data distributions in Fig.~\ref{fig:XibStar2LbK_2MeV}. The measured $\XibStarm$ peak parameters are
\begin{align*}
\delta m_K^{\rm peak} &= 608.3\pm0.8\mev, \\
m(\XibStarm) &= 6227.9\pm0.8\mev, \\
\Gamma(\XibStarm) &= 19.9\pm2.1\mev,
\end{align*}
\noindent where $m(\Lb)=5619.62\pm0.16\pm0.13$\mev~\cite{LHCb-PAPER-2017-011} is used to obtain $m(\XibStarm)$, with signal yields given in Table~\ref{tab:YieldsLb}.
It is notable that the $\Xibm\to\Lb\Km,~\Lb\to\Lc\pim\pip\pim$ signal yield is about 90\% of that of the 
$\Xibm\to\Lb\Km,~\Lb\to\Lc\pim$, even though the initial $\Lb$ sample size is only about 57\% as large. 
This enhancement is expected due to
the higher average $\pt$ of the kaon from the $\Xibm\to\Lb\Km,~\Lb\to\Lc\pim\pip\pim$ decay as compared to the 
$\Xibm\to\Lb\Km,~\Lb\to\Lc\pim$ decay, as discussed previously. 

\subsection{\boldmath{\Xibm mass measurement}}
\begin{figure}[tb]
\centering
\includegraphics[width=0.48\textwidth]{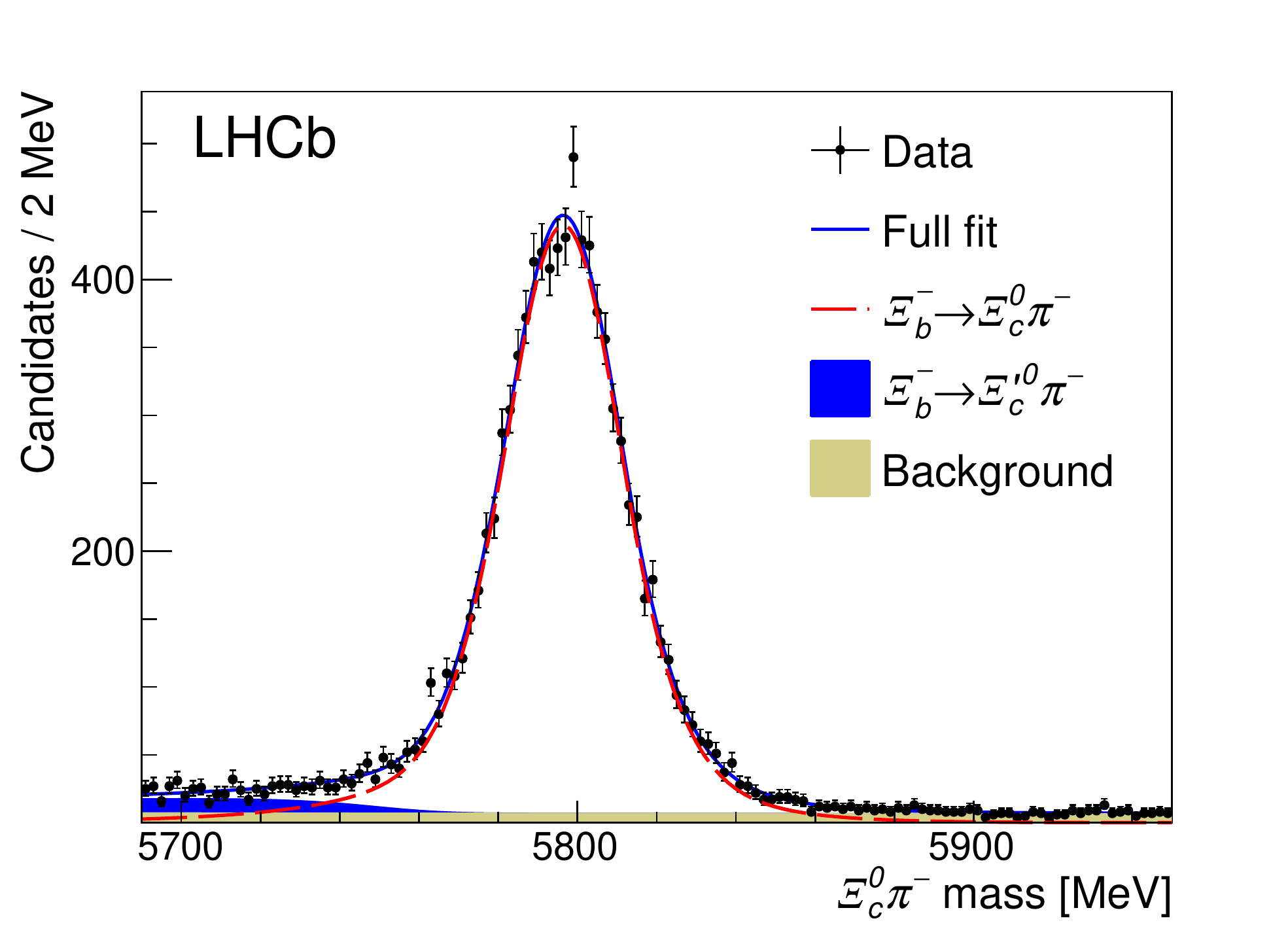}
\includegraphics[width=0.48\textwidth]{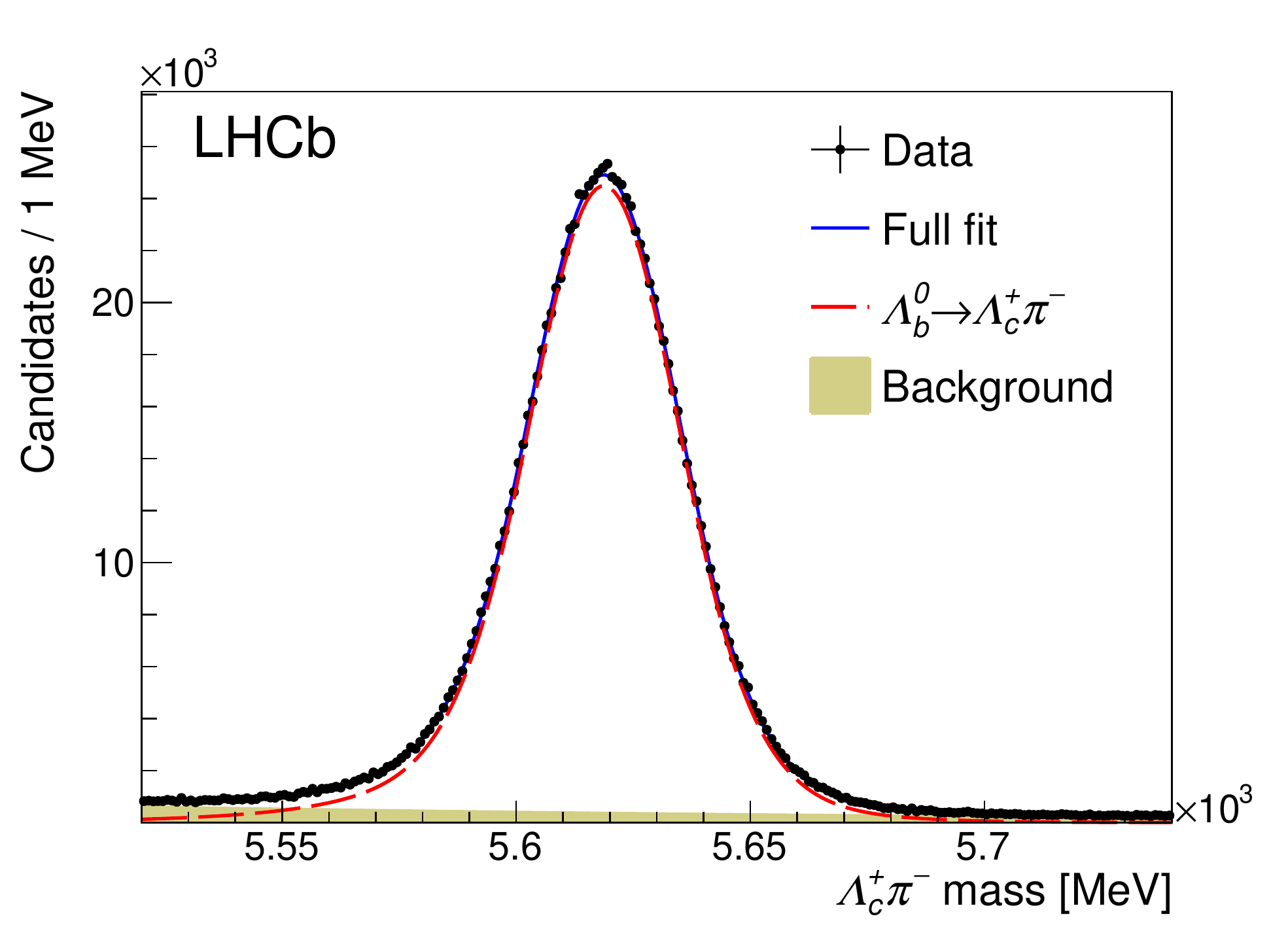}
\caption{\small{Distribution of (left) $\Xicz\pim$ and
(right) $\Lc\pim$ invariant mass for the combined Run 1 and Run 2 data sets, with extra PID selection requirements with respect to the samples shown in Figs.~\ref{fig:Xib} and~\ref{fig:Lb}. The fit projections are overlaid.}}
\label{fig:XibLbMass_forMassMeas_All}
\end{figure}

The large $\Xibm$ and $\Lb$ samples allow for a significant improvement in the uncertainty on  
the $\Xibm$ mass. Only the $H_b\to H_c\pim$ decays are used for this measurement.
The lowest total uncertainty is achieved by measuring the mass difference, $m_{\rm diff} = m_{\rm fit}(\Xibm)-m_{\rm fit}(\Lb)$,
where $m_{\rm fit}(\Xibm)$ and $m_{\rm fit}(\Lb)$ are the peak mass values from fits to the invariant-mass spectra. In $m_{\rm diff}$, the
largest systematic uncertainty, the momentum scale calibration, is greatly reduced. The $\Xibm$ mass is then obtained from $m(\Xibm) = m_{\rm diff} + m(\Lb)$.

All of the previously discussed selection requirements are applied to the samples. Additionally, to render the Cabibbo-suppressed $H_b\to H_c\Km$ 
contribution negligible, a tighter PID requirement is applied to the pion coming directly from the $H_b$ decay. This is done to avoid
the systematic uncertainty associated with the shape and yield of a $H_b\to H_c\Km$ contribution in the mass fit. The efficiency of this
additional selection is 89\% for both the $\Lb$ and $\Xibm$ signal decays.

The binned likelihood fits described previously are applied to the subset of data for this measurement, with the $H_b\to H_c\Km$ background shape
removed. Separate fits are performed on the Run 1 (7 and 8\tev), Run 2 (13\tev) and the full data set. The 
invariant-mass spectra for
$\Xibm$ and $\Lb$ candidates and the fits to the full data sample are shown in Fig.~\ref{fig:XibLbMass_forMassMeas_All}, along
with the full fit and the individual fit components.
The numerical results of the mass fits for each running period and the combined data set are given in Table~\ref{tab:XibMasses}.  
The different values of $m_{\rm fit}(\Lb)$ for Run 1 and Run 2 are a result of the momentum scale uncertainty, which is greatly reduced in $m_{\rm diff}$.
The values of $m_{\rm diff}$ are statistically compatible between the two running periods.

The $\Xibm$ mass is found to be
\begin{align}
m(\Xibm) = 5797.33\pm0.24\mev,
\end{align}
\noindent where the uncertainty is statistical only. 

\begin{table*}[tb]
\begin{center}
\caption{\small{The fitted signal yields and masses of the $\Xibm$ and $\Lb$ peaks and the mass differences,
$m_{\rm diff}\equiv m_{\rm fit}(\Xibm)-m_{\rm fit}(\Lb)$, for each center-of-mass energy and for the full data sample. 
For the last row, the known $\Lb$ mass~\cite{LHCb-PAPER-2017-011} is used.
Uncertainties are statistical only.}}
\begin{tabular}{lccc}
\hline\hline
                     &~~~ Run 1 (7 and 8\tev) & ~~~Run2 (13\tev)       & ~~~All data   \\
\hline
$N(\Xibm)~[10^3]$    & $~~~~\,1.9\pm0.1$   & $~~~~\,7.7\pm0.2$      &  $~~~~\,9.6\pm0.3$ \\
$N(\Lb)~[10^3]$      & $~\,226.7\pm0.7$    & $~\,850.6\pm1.2$       &  $1077.2\pm1.3$ \\
$m_{\rm fit}(\Xibm)~[\mev\,]$   & $5796.12\pm0.57$    & $5796.49\pm0.26$       & $5796.41\pm0.24$ \\
$m_{\rm fit}(\Lb)~[\mev\,]$     & $5618.10\pm0.06$    & $5618.85\pm0.03$       & $5618.70\pm0.03$ \\
\hline
$m_{\rm diff}~[\mev\,]$  & $~\,178.02\pm0.57$  & $~\,177.64\pm0.26$     & $~\,177.71\pm0.24$ \\
\hline
$m(\Xibm)~[\mev\,]$    & $5797.64\pm0.57$    & $5797.26\pm0.26$       & $5797.33\pm0.24$ \\
\hline\hline
\end{tabular}
\label{tab:XibMasses}
\end{center}
\end{table*}

\section{Systematic uncertainties}

Several sources of systematic uncertainty affect the measurements reported in this paper, and are summarized in Table~\ref{tab:syst}.

\subsection{\boldmath{\XibStarz mass and natural width}}
\label{sec:xibstarz_mw}
To estimate the systematic effect of the background shape, three variations on the nominal fit are considered,
including removing the wrong-sign data from the fit, varying the upper range of the mass fit by $\pm$100\mev, and
varying the $\delta M_0$ parameter in the background shape, which was fixed in the nominal fit, by $\pm$10\mev. 
The maximum values among these variations, 0.1\mev for $\delta m_{\pi}^{\rm peak}$ and 1.4\mev for $\Gamma(\XibStarz)$, are 
assigned as systematic uncertainty due to the background shape.

For the signal model, several alternative fits are investigated. Varying the barrier radius between 1\gev$^{-1}$ and 5\gev$^{-1}$, and changing the relativistic Breit--Wigner function to model either an $S$- or $D$-wave decay,
do not change the peak parameters significantly. The peak parameters are found to depend slightly on the assumed mass resolution.
Varying the mass resolution by $\pm$10\% leads to a change in the peak mass and width of 0.1\mev. 
A 0.1\mev uncertainty is assigned to $\delta m_{\pi}^{\rm peak}$ and the $\XibStarz$ width from the signal model. 

The momentum scale calibration uncertainty, known to a precision of $\pm$0.03\%~\cite{LHCb-PAPER-2013-011},
largely cancels in the mass difference. To investigate the effect on $\delta m_{\pi}^{\rm peak}$, the simulation is evaluated with
the momentum scale shifted up and then down by this amount, leading to an uncertainty of 0.2\mev. The energy loss uncertainty
is estimated to be less than 0.1\mev based upon the studies presented in Ref.~\cite{LHCb-PAPER-2011-035}. A 
0.1\mev uncertainty is assigned. 

In computing the uncertainty on
$m(\XibStarz)$, the momentum scale and energy loss are taken to be 
100\% correlated between $\delta m_{\pi}^{\rm peak}$ and $m(\Xibm)$. The total systematic uncertainty is 0.3\mev for
$\delta m_{\pi}^{\rm peak}$, and 0.5\mev and 1.4\mev for the $\XibStarz$ mass and width, respectively.

\subsection{\boldmath{\XibStarm mass and natural width}}

Several variations to the nominal fit are performed to assess the background shape uncertainty. The variations include
changing both the lower (by $+20$\mev) and upper mass limits (by $\pm50$\mev) in the fit. The largest changes in
the peak parameters, 0.4\mev in $\delta m_K^{\rm peak}$ and 1.4\mev in $\Gamma(\XibStarm)$, are assigned as systematic uncertainties.
There is a small excess of events in the $\delta m_K^{\rm peak}$ spectrum in the data near 520\mev. In an alternative fit, a second peak is 
included in the fit model for both mass spectra. The second peak is found to be statistically insignificant, however, its inclusion changes the $\XibStarm$ 
mass by 0.1\mev and its width by 0.8\mev. These values are added in quadrature with the values found from varying the fit range to 
arrive at a background systematic uncertainty of 0.4\mev and 1.5\mev on $\delta m_K^{\rm peak}$ and $\Gamma(\XibStarm)$, respectively.

For the signal model uncertainty, a similar set of variations is carried out as for the $\XibStarz$ case, and only the
width shows any sensitivity to the $\pm$10\% variation in the mass resolution. The change of 0.1\mev is assigned
as an uncertainty on the $\XibStarm$ width.

The momentum and energy scale uncertainties each lead to a 0.1\mev uncertainty on \mbox{$\delta m_{K}^{\rm peak}$}. In combining \mbox{$\delta m_{K}^{\rm peak}=608.3\pm0.8\pm0.4$\mev} with $m(\Lb)$~\cite{LHCb-PAPER-2017-011} to obtain $m(\XibStarm)$, the momentum scale and
energy loss portion of the systematic uncertainties are taken to be 100\% correlated. The resulting systematic uncertainty on $m(\XibStarm)$ is 0.5\mev. 

\subsection{\boldmath{Production ratio $R(\Xibm\pip)$}}

In the measurement of $R(\Xibm\pip)$, the sources of uncertainty include the
signal and background shapes in the $\Xibm$ and $\XibStarz$ mass fits, and the relative efficiency estimate.
For the $\Xibm$ mass fit, the signal yield is evaluated with an alternative signal model comprised of the
sum of two Gaussian functions, where the means need not be the same and the widths are allowed to vary in the fit.
The yield in this alternative fit changes by 2\%, which is taken as a systematic error.
The uncertainty due to the background shape is studied by changing to a
Chebyshev polynomial, which leads to a 1.4\% change in the yield. The upper end of the mass fit is reduced from 5950\mev to 5900\mev,
and the 0.4\% change in signal yield is assigned as systematic uncertainty.
These two contributions are added in quadrature, resulting in an uncertainty of 1.5\% due to the $\Xibm$ background shape.

Variations in the $\XibStarz$ background shape are also considered for the uncertainty on $R(\Xibm\pip)$. The same set of 
variations that were performed for the $\XibStarz$ mass and width are considered. Adding the changes in 
yield in quadrature leads to a 5.6\% uncertainty due to the $\XibStarz$ background shape. Several variations in the signal model are
considered, and the only non-negligible change in signal yield occurs when
a non-relativistic Breit--Wigner function is used in place of the relativistic Breit--Wigner shape. The 0.7\% change in the signal yield
is assigned as an uncertainty to the $\XibStarz$ yield.

The relative efficiency depends on the extent to which the simulation properly models the $(\pt,\eta)$ spectrum of
$\Xibm$ and the $\XibStarz$ production spectra. The large $\Xibm$ sample allows for a precise tuning of the $\kappa$ parameters, 
so that the $\pt$ and $\eta$ spectrum in simulation is well matched to that of the data. Due to the low signal yields
in the $\XibStarz$ sample, it is estimated that the $\kappa$ parameters have an uncertainty of $\pm$0.005 units. A larger shift than 0.005 units leaves the simulation in clear disagreement with the background-subtracted data. Varying the
$\kappa_T$ parameter by this amount leads to an 8\% change in $\epsilon_{\rm rel}$. This change is due almost entirely
to the $\pt>700$\mev requirement on the $\pip$ meson in the $\XibStarz$ decay. A $\pm0.005$ unit variation in $\kappa_z$ is 
also investigated, but leads to a negligible change in the relative efficiency. The $\pip$ tracking efficiency correction
has an uncertainty of 2.1\%, which includes a 1.5\% contribution from the calibration using $\jpsi\to\mup\mun$ decays and 1.4\% due to the difference in material interactions between muons and pions~\cite{LHCb-DP-2013-002}. The finite simulated sample sizes lead to an additional systematic uncertainty of 1.2\%.

\subsection{\boldmath{\Xibm mass}}
The systematic uncertainty in $m_{\rm diff}$ is studied by performing alternative fits to the data, and assigning the
change in $m_{\rm diff}$ with respect to the nominal value as a systematic uncertainty.
The background shape uncertainty is estimated by using a Chebychev polynomial instead of the
exponential background shape (0.05\mev), reducing the upper limit of the fit range by 50\mev (0.06\mev), and
fitting with a finer binning (0.02\mev). The total background shape uncertainty is taken as the quadrature sum,
which is 0.08\mev. The signal shape uncertainty is assigned by changing the way the tail parameters are treated in
the signal function. For the $\Xibm$ mass fit, they are changed from fixed values to floating values,
and for the $\Lb$, they are changed from floating values to fixed values based on the simulation. These variations lead to a change in
$m_{\rm diff}$ of 0.10\mev, which is assigned as the signal shape uncertainty. The momentum scale and 
energy loss uncertainties are unchanged from the previous result~\cite{LHCb-PAPER-2014-048}, and are
0.08\mev and 0.06\mev, respectively. Adding these uncertainties in quadrature, the total uncertainty
on $m_{\rm diff}$ is 0.16\mev. 

In combining \mbox{$m_{\rm diff}=177.71\pm0.24\pm0.16\mev$} with $m(\Lb)$~\cite{LHCb-PAPER-2017-011} to obtain $m(\Xibm)$, the momentum scale and
energy loss portion of the systematic uncertainties are taken to be 100\% correlated. The remainder of the uncertainties are taken to be uncorrelated. The resulting systematic uncertainty on $m(\Xibm)$ is 0.29\mev. 

\begin{table*}[bt]
\begin{center}
\setlength\extrarowheight{2pt}
\caption{\small{Summary of systematic uncertainties on quantities related to the $\XibStarz$
($\dmPiPeak$, $\Gamma(\XibStarz)$, $R(\Xibm\pip)$), the $\XibStarm$ ($\dmKPeak$, $\Gamma(\XibStarm$),
and the $\Xibm$ mass ($m_{\rm diff}$) measurements. The statistical uncertainties are also reported for comparison.}}
\begin{tabular*}{\textwidth}{l@{\extracolsep{\fill}}*{6}{c}}
\hline\hline
Source                  & \multicolumn{3}{c}{$\XibStarz$}  & \multicolumn{2}{c}{$\XibStarm$} & $\Xibm$ \\   
\cline{2-4}
\cline{5-6}
\cline{7-7}
                        & $\dmPiPeak$ &  $\Gamma$ & $R(\Xibm\pip)$ &  $\dmKPeak$ &  $\Gamma$ & $m_{\rm diff}$ \\
                        &  $[\mev]$     & $[\mev]$    &  [\%]          &  $[\mev]$     &    $[\mev]$ &     $[\mev]$   \\ 
\hline         
$\XibStarz$ back. shape   &  0.1    &     1.4   &  5.6   &   -    &      -      &   - \\
$\XibStarz$ signal shape &  0.1    &     0.1   &  0.7   &   -    &      -      &   - \\
$\XibStarm$ back. shape   &   -     &      -    &   -    &  0.4   &     1.5     &   - \\
$\XibStarm$ signal shape &   -     &      -    &   -    &  0.0   &     0.1     &   - \\
$\Xibm,\,\Lb$ back. shape &   -     &      -    &  1.5   &   -    &      -      & 0.08 \\
$\Xibm,\,\Lb$ signal shape&   -    &      -    &  2.0   &  -     &     -       & 0.10 \\
Momentum scale           & 0.2     &    0.0    &   -    &  0.1   &     0.0     & 0.08 \\
Energy loss              & 0.1     &    0.0    &   -    &  0.1   &     0.0     & 0.06 \\
Production spectra       &   -     &      -    &  8.0   &  -     &     -       &  -    \\ 
$\pip$ tracking efficiency&  -     &      -    &  2.1   &  -     &     -       &  -    \\
Simulated sample size    &   -     &      -    &  1.2   &  -     &     -       &  -  \\
\hline
Total systematic  &      0.3       &     1.4   &  10.4  &  0.4   &     1.5      & 0.16 \\ 
\hline
\\[-3ex]
Statistical       &  $^{+1.4}_{-1.5}$ & $^{+5.0}_{-4.1}$ & 18 &  0.8   &     2.1          & 0.24     \\
\\[-3ex]
\hline\hline
\end{tabular*}
\label{tab:syst}
\end{center}
\end{table*}

\section{Summary}
Using $pp$ collision data at $\sqrt{s}=7,~8$ and $13$\tev, corresponding to an integrated luminosity of 8.5\invfb, a new $\Xibz$ baryon, referred to as $\XibStarz$, is reported with a statistical significance of 10\,$\sigma$. The mass difference, mass and natural width of the peak 
are measured to be
\begin{align*}
\delta m_{\pi}^{\rm peak} &= 429.8^{\,+1.4}_{\,-1.5}\pm0.3\mev, \\
m(\XibStarz) &= 6227.1^{\,+1.4}_{\,-1.5}\pm0.5\mev, \\
\Gamma(\XibStarz) &= 18.6^{\,+5.0}_{\,-4.1}\pm1.4\mev, 
\end{align*}
\noindent where the first uncertainty is statistical and the second is experimental systematic. 

The relative production rate of the $\XibStarz$ state at $\sqrt{s}=13$\tev is 
measured through its decay to $\Xibm\pip$ to be
\begin{align*}
R(\Xibm\pip) \equiv \frac{f_{\XibStarz}}{f_{\Xibm}}\BF(\XibStarz\to\Xibm\pip) = 0.045\pm0.008\pm0.004.
\end{align*}
\noindent This is consistent with the values of $R(\Xibz\pim)$ found in Ref.~\cite{LHCb-PAPER-2018-013} for the
$\XibStarm$ state. The value of $R(\Xibm\pip)$ can also be compared to the corresponding value found for the
lower-mass $\XibStarZero$ state of $0.28\pm0.03\pm0.01$~\cite{LHCb-PAPER-2016-010}. Additional unobserved decay
modes, such as $\XibStarZero\to\Xibz\piz$ and $\XibStarz\to(\Xibz\piz,\Lb\Kzb)$, would clearly contribute to the total
production rate of these excited states, but are yet to be observed.

From a sample of $\XibStarm\to\Lb\Km$ signal decays that is approximately four times 
larger than that which was used in the first observation of the $\XibStarm$ baryon~\cite{LHCb-PAPER-2018-013}, an updated 
measurement of the $\XibStarm$ mass and natural width is presented. The values obtained are
\begin{align*}
\delta m_{K}^{\rm peak} &= 608.3\pm0.8\pm0.4\mev, \\
m(\XibStarm) &= 6227.9\pm0.8\pm0.5\mev,\\
\Gamma(\XibStarm) &= 19.9\pm2.1\pm1.5\mev, 
\end{align*}
\noindent which supersede the results in Ref.~\cite{LHCb-PAPER-2018-013}. The measured masses of the $\XibStarz$ and
$\XibStarm$ states are consistent with them being isospin partners.

Lastly, from a sample of about 10\,000 $\Xibm\to\Xicz\pim$ and 1 million $\Lb\to\Lc\pim$ signal decays, the mass difference 
between the two $b$ baryons and the $\Xibm$ mass are measured to be
\begin{align*}
m_{\rm diff} &= 177.71\pm0.24\pm0.16\mev, \\
m(\Xibm) &= 5797.33\pm0.24\pm0.29\mev.
\label{eq:XibMass}
\end{align*}
\noindent The result obtained here represents the single most precise determination of the $\Xibm$ mass. It is consistent with previous measurements and is about a factor of 1.6 times more precise than the current world average~\cite{PDG2020}, and it supersedes the measurement reported in Ref.~\cite{LHCb-PAPER-2014-048}.

With the current data sample, it cannot be excluded that there are two or more narrower, closely spaced states 
contained within the peaks referred to as $\XibStarm$ and $\XibStarz$.
With larger data samples in the future, it should be possible to probe whether these peaks are comprised of
narrower states.

\section*{Acknowledgements}
%
% These Acknowledgements valid from 3-May-2019
%
\noindent We express our gratitude to our colleagues in the CERN
accelerator departments for the excellent performance of the LHC. We
thank the technical and administrative staff at the LHCb
institutes.
We acknowledge support from CERN and from the national agencies:
CAPES, CNPq, FAPERJ and FINEP (Brazil); 
MOST and NSFC (China); 
CNRS/IN2P3 (France); 
BMBF, DFG and MPG (Germany); 
INFN (Italy); 
NWO (Netherlands); 
MNiSW and NCN (Poland); 
MEN/IFA (Romania); 
MSHE (Russia); 
MICINN (Spain); 
SNSF and SER (Switzerland); 
NASU (Ukraine); 
STFC (United Kingdom); 
DOE NP and NSF (USA).
We acknowledge the computing resources that are provided by CERN, IN2P3
(France), KIT and DESY (Germany), INFN (Italy), SURF (Netherlands),
PIC (Spain), GridPP (United Kingdom), RRCKI and Yandex
LLC (Russia), CSCS (Switzerland), IFIN-HH (Romania), CBPF (Brazil),
PL-GRID (Poland) and OSC (USA).
We are indebted to the communities behind the multiple open-source
software packages on which we depend.
Individual groups or members have received support from
AvH Foundation (Germany);
EPLANET, Marie Sk\l{}odowska-Curie Actions and ERC (European Union);
A*MIDEX, ANR, Labex P2IO and OCEVU, and R\'{e}gion Auvergne-Rh\^{o}ne-Alpes (France);
Key Research Program of Frontier Sciences of CAS, CAS PIFI,
Thousand Talents Program, and Sci. \& Tech. Program of Guangzhou (China);
RFBR, RSF and Yandex LLC (Russia);
GVA, XuntaGal and GENCAT (Spain);
the Royal Society
and the Leverhulme Trust (United Kingdom).

\clearpage
\newpage
\addcontentsline{toc}{section}{References}
%\setboolean{inbibliography}{true}
\bibliographystyle{LHCb}
\bibliography{main,standard,LHCb-PAPER,LHCb-CONF,LHCb-DP,LHCb-TDR}

\newpage
% LHCb collaboration author list
% Data extracted on October 22nd, 2020 at 7:52pm for reference date 03-Sep-2020
\centerline
{\large\bf LHCb collaboration}
\begin
{flushleft}
\small
R.~Aaij$^{31}$,
C.~Abell{\'a}n~Beteta$^{49}$,
T.~Ackernley$^{59}$,
B.~Adeva$^{45}$,
M.~Adinolfi$^{53}$,
H.~Afsharnia$^{9}$,
C.A.~Aidala$^{84}$,
S.~Aiola$^{25}$,
Z.~Ajaltouni$^{9}$,
S.~Akar$^{64}$
% ...

{\footnotesize \it

% ...
$ ^{9}$Universit{\'e} Clermont Auvergne, CNRS/IN2P3, LPC, Clermont-Ferrand, France\\
% ...
$ ^{25}$INFN Sezione di Milano, Milano, Italy\\
% ...
$ ^{31}$Nikhef National Institute for Subatomic Physics, Amsterdam, Netherlands\\
% ...
$ ^{45}$Instituto Galego de F{\'\i}sica de Altas Enerx{\'\i}as (IGFAE), Universidade de Santiago de Compostela, Santiago de Compostela, Spain\\
% ...
$ ^{49}$Physik-Institut, Universit{\"a}t Z{\"u}rich, Z{\"u}rich, Switzerland\\
% ...
$ ^{53}$H.H. Wills Physics Laboratory, University of Bristol, Bristol, United Kingdom\\
% ...
$ ^{59}$Oliver Lodge Laboratory, University of Liverpool, Liverpool, United Kingdom\\
% ...
$ ^{64}$University of Cincinnati, Cincinnati, OH, United States\\
% ...
$ ^{67}$Syracuse University, Syracuse, NY, United States\\
% ...
$ ^{84}$University of Michigan, Ann Arbor, United States, associated to $^{67}$\\
\bigskip
% ...
\medskip
}
\end{flushleft}

\end{document}